\documentclass[a4paper,12pt]{article}
\usepackage{amssymb, amsmath, amsthm, mathrsfs, latexsym, mathtools,authblk}
\usepackage[flushleft]{threeparttable}
\usepackage{setspace, lscape, comment}
\usepackage[font=normal]{subfig}
\usepackage{fnpct}
\usepackage[longnamesfirst, round]{natbib}
\newtheorem{assumption}{Assumption}
\numberwithin{assumption}{section}

\newtheorem{theorem}{Theorem}

\theoremstyle{definition}
\newtheorem{remark}{Remark}

\usepackage[margin=1in]{geometry}
\allowdisplaybreaks[1]
\usepackage[colorlinks=true, linkcolor=blue, filecolor=blue, urlcolor=blue, citecolor=blue, driverfallback=dvipdfmx]{hyperref}
\usepackage[pdflatex]{graphicx,psfrag}

\title{Inference on Local Average Treatment Effects for Misclassified Treatment}
\author{Takahide Yanagi\footnote{Graduate School of Economics, Hitotsubashi University,
		2-1 Naka, Kunitachi, Tokyo 186-8601, Japan.
		Email: \href{mailto:t.yanagi@r.hit-u.ac.jp}{t.yanagi@r.hit-u.ac.jp}
	}
}
\date{This version: April 10, 2018 $\qquad$ First version:  January 27, 2017}

\begin{document}

\maketitle

\begin{abstract}
We develop point-identification for the local average treatment effect when the binary treatment contains a measurement error.
The standard instrumental variable estimator is inconsistent for the parameter since the measurement error is non-classical by construction.
We correct the problem by identifying the distribution of the measurement error based on the use of an exogenous variable that can even be a binary covariate.
The moment conditions derived from the identification lead to generalized method of moments estimation with asymptotically valid inferences. 
Monte Carlo simulations and an empirical illustration demonstrate the usefulness of the proposed procedure.
\bigskip
	
\bigskip
	
\end{abstract}

\newpage

\section{Introduction}\label{sec:intro}
The local average treatment effect (LATE) is a popular causal parameter in the microeconometric literature (e.g., \citealp[Chapter 4]{angrist2008mostly}).
It represents the average causal effect of a binary endogenous treatment $T^*$ on an outcome $Y$ for the unit whose treatment status changes depending on the value of a binary instrument $Z$.
\cite{imbens1994identification} show that the instrumental variable (IV) estimator may identify the LATE.
While the identification requires the precise measurement of the {\it true} binary treatment $T^*$ in addition to identification conditions, in practice, the observed binary treatment $T$ may be mismeasured for $T^*$.
LATE applications may involve such a degree of misclassification that the actually treated unit can even be misrecorded as untreated and vice versa.
While we are not aware of the presence of measurement errors in practice, ignoring such errors may lead to drawing misleading empirical implications.

As an application, let us consider the causal analysis of returns to schooling, one of the most important applications of LATE inferences.
In this case, $Y$ is an individual outcome such as wages, $T^*$ is an indicator of educational attainment such as a college degree, and $Z$ is a variable indicating whether the individual lives close to a college (e.g., \citealp{card1993using}).
Educational attainment may be mismeasured because of a random recording error or the provision of intentionally/unintentionally false statements.
Indeed, several studies have pointed out the prevalence of misreported schooling (\citealp{kane1995labor}; \citealp{card1999causal}; \citealp{black2003measurement}; \citealp{battistin2014misreported}).

The present study contributes to the econometric literature by proposing a novel inference procedure for the LATE with the mismeasured treatment.
The measurement error brings out a bias such that the conventional IV estimator under- or overestimates the LATE.
While the IV estimation solves the problem of the {\it classical} measurement error that is uncorrelated with the true variable (e.g., \citealp[Chapters 4 and 5]{wooldridge2010econometric}), the bias for the LATE is caused in our situation since the measurement error for the binary variable must be {\it non-classical}.
That is, the error is correlated with the true unobserved treatment, because the support of the measurement error for the discrete variable depends on the true variable (e.g., \citealp{aigner1973regression}).
The bias for the LATE depends on the distribution of the measurement error, meaning a failure to identify the LATE.

\begin{figure}[!h]
    \begin{center}
        \subfloat[Covariate $V$]{
            \includegraphics[width=80mm, bb=0 0 117 109]{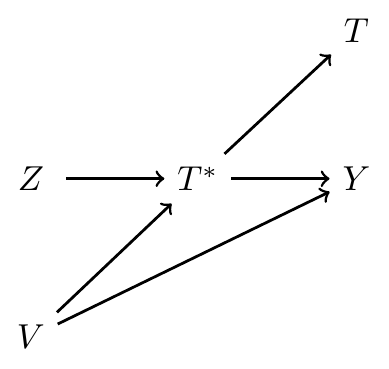}
        }
        \subfloat[Instrument $V$]{
            \includegraphics[width=80mm, bb=0 0 117 109]{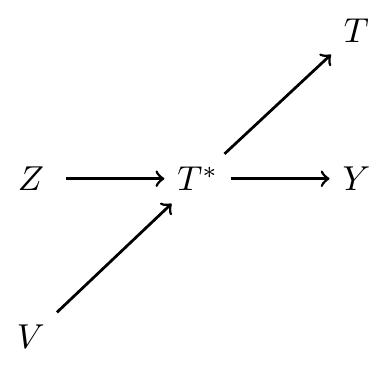}
        }
        \newline
        \subfloat[Repeated measure $V$]{
        	\includegraphics[width=80mm, bb=0 0 117 109]{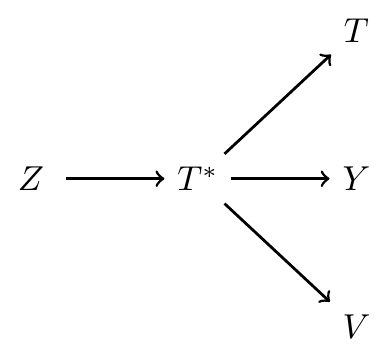}
        }
        \caption{Each panel indicates the intuitive relationship between the outcome $Y$, true unobserved treatment $T^*$, mismeasured observed treatment $T$, instrument $Z$, and exogenous variable $V$ in each situation.
        Each arrow indicates that the left variable affects the right variable.}
        \label{fig:exogenous}
    \end{center}    
\end{figure}

To correct the identification bias, we develop point-identification for the LATE and the distribution of the measurement error based on the availability of an exogenous observable variable, say $V$.
The exogenous variable $V$ can be a covariate, instrument, or repeated measure of the treatment as long as $V$ satisfies our identification conditions.
Intuitively, Figure \ref{fig:exogenous} illustrates the relationship between such exogenous variables and other variables.
Many previous studies have corrected problems due to measurement errors by using such exogenous variables (e.g., \citealp{hausman1991identification}; \citealp{lewbel1997constructing,lewbel1998semiparametric}; \citealp{schennach2007instrumental}; \citealp{hu2008identification}; \citealp{hu2008instrumental}), and our identification builds on this strand of the literature.
In particular, we extend the results by \citet{mahajan2006identification} and \citet{lewbel2007estimation} on the mismeasured exogenous treatment to the LATE inference with the mismeasured endogenous treatment.

The main idea behind our identification is, under a set of empirically plausible conditions, to derive moment conditions no fewer than the parameters including the LATE, true first-stage regression, and distribution of the measurement error.
We introduce three key conditions when $V$ is non-binary, and we require an additional condition when $V$ is binary.
Firstly, the measurement error of the treatment is {\it non-differential} in the sense that mismeasured $T$ does not affect the mean outcome once true $T^*$ is conditioned on.
Secondly, $V$ has to satisfy a relevance condition that requires that $V$ relates to $T^*$.
Thirdly, $V$ has to satisfy an exclusion restriction under which $V$ cannot affect the difference in the conditional means of $Y$ and the distribution of the measurement error of $T$.
Finally, when $V$ is binary, we need the additional condition under which the distribution of the measurement error does not depend on $Z$.
These conditions may be satisfied when $V$ is a covariate, instrument, or repeated measure of the treatment.
Importantly, our exclusion restriction does not rule out the possibility that $V$ directly affects $Y$ and does hold if $V$ affects the outcomes with and without the true treatment equally.

The moment conditions derived from the identification lead to moment-based estimators for the parameters.
We propose adopting \citeauthor{hansen1982large}'s (\citeyear{hansen1982large}) generalized method of moments (GMM) estimator because of its popularity in the literature.
Desirably, the GMM inference is easy to implement in practice, and its asymptotic properties are well understood.
The asymptotically valid inferences can be developed in the usual manner.\footnote{An {\ttfamily R} package to implement the proposed GMM inference is available on the author's website.}

We illustrate the usefulness of the proposed GMM inference based on Monte Carlo simulations and an empirical illustration.
The simulations show that our GMM inference works well in finite samples.
The empirical illustration estimates the returns to schooling and the distribution of misreported schooling using \citeauthor{card1993using}'s (\citeyear{card1993using}) data by utilizing two indicators of college proximity as exogenous variations $Z$ and $V$.\footnote{While our empirical illustration utilizes two indicators of college proximity, the literature on the returns to schooling suggests many other potential exogenous variables such as the quarters of birth (\citealp{angrist1991does}), parental/sibling education (\citealp{altonji1996effects}), and the sex of siblings (\citealp{butcher1994effect}).
	See, for example, \citet{card1999causal, card2001estimating} for a survey on returns to schooling analyses based on such exogenous variables.
	Such exogenous variables are candidates for exogenous variables to overcome endogeneity and measurement errors simultaneously in our proposed inference.
}
Our relevance condition requires that whether living close to colleges relates to schooling, and we demonstrate its validity.
Our exclusion restriction allows college proximity to relate to the factors affecting wages and seems more plausible than the exclusion restriction for the standard IV inference.
We find that the conventional inference overestimates the LATE because of misreported education and our GMM inference corrects the problem successfully.

\paragraph{Paper organization.}
Section \ref{sec:literature} reviews related studies.
Section \ref{sec:setting} introduces the setup, reviews the LATE inference, and explains the problem due to the measurement error.
Section \ref{sec:identification} develops the identification.
Sections \ref{sec:monte} and \ref{sec:empirical} present the simulations and empirical illustration, respectively.
Section \ref{sec:conclusion} concludes.
Appendix \ref{sec:proof} gives the proofs of the theorems.
The supplementary appendix contains auxiliary results.

\section{Related literature} \label{sec:literature}

The present study builds on the literature on misclassification problems (i.e., problems due to measurement errors for discrete variables) on which many studies contribute.
For example, \citet{aigner1973regression}, \citet{bollinger1996bounding}, \citet{hausman1998misclassification}, \citet{frazis2003estimating}, \cite{molinari2008partial}, \cite{hu2008identification}, \cite{imai2010causal}, and \citet{ura2016instrumental} consider linear and non-linear models with possibly misclassified discrete variables.
Since misclassification causes a non-classical measurement error that is correlated with the true variable, it fails to identify the parameter of interest.
Since the identification bias due to misclassification depends on the misclassification probabilities (i.e., the probabilities that the truly treated is misclassified as the untreated and vice versa), the literature has developed how to infer the misclassification probabilities.

In particular, our identification of the misclassification probabilities relates to the results presented by \citet{mahajan2006identification} and \citet{lewbel2007estimation} among the studies in the misclassification literature.
Both works identify and estimate the conditional mean of the outcome $Y$ given the binary exogenous treatment $T^*$ and its difference $E(Y|T^*=1) - E(Y|T^*=0)$ based on the availability of the misclassified binary treatment $T$ and an exogenous variable $V$.\footnote{While \citet{mahajan2006identification} and \citet{lewbel2007estimation} allow for the presence of other control variables, we omit such variables here for perspicuity.
}
However, their identification conditions have important distinctions for the relationship between $Y$ and $V$ and for the number of the elements that $V$ takes.
On the one hand, \citet{mahajan2006identification} requires that $E(Y|T^*, V) = E(Y|T^*)$ so that $V$ cannot be a covariate in the sense that $V$ cannot directly affect $Y$, although he allows $V$ to be binary.
On the other hand, \citet{lewbel2007estimation} allows $V$ to be a covariate by introducing an exclusion restriction under which $E(Y|T^*=1, V) - E(Y|T^* = 0, V)$ does not depend on $V$, but he requires that $V$ has to take at least three values, meaning that $V$ cannot be binary.

Our identification extends the results presented by \citet{mahajan2006identification} and \citet{lewbel2007estimation} by allowing $V$ to be binary and a covariate simultaneously.
These features of $V$ are empirically important since we are rarely confident about whether an exogenous variable affects the outcome and often observe only a binary exogenous variable.
Our identification allows for both these features by utilizing the condition that the instrument $Z$ is not related to the misclassification probabilities, instead of excluding the direct effect of $V$ on $Y$ and binariness of $V$.
This condition for the misclassification probabilities could be empirically plausible since $Z$ should be assigned as good as randomly in the LATE inference (see \citealp{angrist2008mostly}).
Indeed, some of the studies of LATE inferences with a mismeasured treatment discussed below utilize the same condition to identify a parameter similar to the LATE or to partially identify the LATE.
We also stress that we cannot use the estimation procedures of \citet{mahajan2006identification} and \citet{lewbel2007estimation} to infer the LATE since they intend to infer the conditional mean of $Y$ given $T^*$ and its difference.

\citet{mahajan2006identification} also discusses identification when the true binary treatment is endogenous and misclassified as an extension of his main result, although the recent study by \citet{ditraglia2015mis} points out that \citeauthor{mahajan2006identification}'s (\citeyear{mahajan2006identification}) result fails.
The intuitive reason for this failure is that it is impossible for his approach to correct both endogeneity and misclassification by using solely one instrument.
See \citet{ditraglia2015mis} for details on this problem.
We note that such a problem does not occur in our proposed inference since we utilize two exogenous variations, $Z$ and $V$.

To the best of our knowledge, LATE inferences with the binary, endogenous, and misclassified treatment have been examined by \citet{battistin2014misreported}, \citet{calvi2017women}, \citet{ditraglia2015mis}, and \citet{ura2017hetero}.

\cite{battistin2014misreported} use two repeated measures for the treatment in addition to the observed treatment $T$ to develop point-identification and a semiparametric estimation for the LATE in a mixture model.
Their necessity is the presence of multiple repeated measures for the true treatment from resurvey data.
Unlike their approach, our proposed inference may be applicable even when one may utilize a covariate.

\citet{calvi2017women} utilize a repeated measure of the treatment in addition to the observed treatment $T$ to identify a causal parameter similar to the standard LATE, which they call the mismeasurement robust LATE (MR-LATE).
While the MR-LATE can be identified under less restrictive assumptions, it is not identical to the standard LATE in general.
We stress that the MR-LATE is different from the standard LATE under the identification conditions developed in the present study.

\cite{ditraglia2015mis} develop partial- and point-identification for the LATE in a non-parametric additively separable model with moment restrictions on the measurement error and an additively separable error.
They do not require the availability of an exogenous variable and/or resurvey data, but demand that the individual treatment effect does not depend on the unobservables.
On the contrary, we allow arbitrary heterogeneous treatment effects, although we still need an exogenous variable.

\cite{ura2017hetero} proposes sharp partial-identification for the LATE with the mismeasured treatment in a general setting.
His approach allows for an endogenous measurement error and does not require resurvey data and/or the use of an exogenous variable, but his result does not attain point-identification for the LATE.

Many econometric studies examine non-classical measurement errors and/or non-linear errors-in-variables models for other settings, and our proposed inference also builds on this body of the literature.
For example, \citet{amemiya1985instrumental}, \citet{hsiao1989consistent}, \citet{horowitz1995identification}, \citet{hu2008instrumental}, \citet{hu2015closed}, and \cite{song2015estimating} study such measurement error models for microeconometric applications.
See \citet{bound2001measurement}, \citet{chen2011nonlinear}, and \citet{schennach2016recent} for excellent reviews of the literature in this regard.

The present study also builds on recent research using covariates to solve endogeneity or measurement errors.
In particular, our exclusion restriction and relevance condition relate to conditional covariance restrictions given covariates in \citet{caetano2017identifying}.
We compare their restrictions with ours in Remark \ref{remark:covariance}, and we here stress that both papers have different aims.
They focus on estimating marginal effects in general IV models without measurement errors, while our aim is to infer the LATE with misclassification.
Building on the important result by \citet{caetano2017identifying}, for example, \citet{ben2017identification} also utilize analogous restrictions for covariates to solve measurement errors.
However, their setting is also different from ours.

\section{Setting}\label{sec:setting}
This section explains the setting considered in this study.
Section \ref{sec:data} introduces the econometric model and briefly reviews the LATE inference without a measurement error.
Section \ref{sec:problem} discusses the problem due to a measurement error for the treatment.

\subsection{The model and the LATE}\label{sec:data}
Let $(\Omega, \mathcal{F}, P)$ be the common probability space where the random variables introduced below are defined.
We have a random sample of $(Y, T, Z, V)$, where $Y \in \mathbb{R}$ is an outcome, $T \in\{0,1\}$ is a possibly mismeasured treatment, $Z \in\{0,1\}$ is an IV, and $V \in supp(V) \subset \mathbb{R}$ is an exogenous variable such as a covariate, instrument, or repeated measure of the treatment.
The true unobservable treatment $T^* \in \{0,1\}$ may be endogenous because of omitted variables in the sense that some unobservables relate to both $Y$ and $T^*$.
The observed $T$ may contain a measurement error, meaning that $T \neq T^*$ in general.

The availability of an exogenous variable $V$ is essential for our procedure.
While the standard LATE inference does not require such exogenous variables, our analysis needs $V$ to correct the misclassification problem.
There are many empirical situations in which we can utilize such exogenous variables.
For example, when survey data with rich information are available, we may utilize covariates and/or instruments affecting $T^*$ and/or $Y$ as $V$.
Further, when we access survey and resurvey data, we may observe a repeated binary measure $V$ for true $T^*$ in addition to $T$.
Note that in this case $V$ might also contain a measurement error for $T^*$ and $V$ is not identical to $T$ in general.
Section \ref{sec:identification} discusses that $V$ has to satisfy some conditions such as exclusion restrictions and a relevance condition to identify the LATE and the distribution of the measurement error.

For example, in the returns to schooling analysis, $Y$ is wages, $T^*$ is a college degree, $Z$ is the IV indicating whether the individual lives close to a four-year college, and $V$ is a covariate or instrument (e.g., two-year college proximity or the quarter of birth), or a repeated measure of the degree.

The aim of the inference is to examine the causal relationship between $T^*$ and $Y$.
Let $Y_0$ and $Y_1$ be the potential outcomes when the unit is truly untreated ($T^*=0$) and when it is treated ($T^*=1$), respectively.
Similarly, let $T^*_0$ and $T^*_1$ be the potential true treatment statuses when $Z=0$ and $Z=1$, respectively.
We can write $Y= Y_0 + T^* (Y_1-Y_0)$ and $T^* =T_0^* + Z (T_1^* - T_0^*)$.
The individual causal effect of $T^*$ on $Y$ is $Y_1 - Y_0$, which may be heterogeneous across units depending on the observables and/or unobservables.

We define the following subsets of the common probability space $(\Omega, \mathcal{F}, P)$.
\begin{equation*}
\begin{array}{ll}
	\text{Always taker: } \quad & A \coloneqq \{\omega \in \Omega: T_0^*(\omega)=1, \; T_1^*(\omega)=1  \},\\
	\text{Complier:} & C \coloneqq \{\omega \in \Omega: T_0^*(\omega)=0, \; T_1^*(\omega)=1  \},\\
	\text{Defier:} & D \coloneqq \{\omega \in \Omega: T_0^*(\omega)=1, \; T_1^*(\omega)=0  \},\\
	\text{Never taker:} & N \coloneqq \{\omega \in \Omega: T_0^*(\omega)=0, \; T_1^*(\omega)=0  \}.
\end{array}
\end{equation*}
Intuitively, $A$ or $N$ is the set of units that always take or deny, respectively, the treatment (in the sense of true $T^*$), $C$ is that of units whose treatment statuses are positively affected by $Z$, and $D$ is that of units whose treatment statuses are negatively affected by $Z$.

The parameter of interest is the LATE, which is defined as
\begin{align}\label{eq:parameter}
	E(Y_1-Y_0|C)=E(Y_1-Y_0|T_1^* > T_0^*).
\end{align}
Without the measurement error, \citet[Theorem 1]{imbens1994identification} show that the LATE is identified by the IV estimand based on $(Y,T^*,Z)$:
\begin{align}\label{eq:LATE}
	E(Y_1-Y_0|C) = \beta^*,
\end{align}
where $\beta^* = \Delta \mu / \Delta p^*$ is the IV estimand with $\Delta \mu \coloneqq \mu_1 - \mu_0$, $\Delta p^* \coloneqq p_1^* - p_0^*$, $\mu_z \coloneqq E(Y|Z=z)$, and $p_z^* \coloneqq E(T^*|Z=z) = \Pr(T^*=1|Z=z)$.
Note that, if we observe $(Y,T^*,Z)$, we can consistently estimate $\beta^*$ by the two-stage least squares estimation.

Formally, we need the following conditions to establish the identification in \eqref{eq:LATE}.
These are essentially the same as the conditions in \citet{imbens1994identification}.

\begin{assumption}\label{as:late1}
	$(Y_1, Y_0, T_1^*, T_0^*)$ is independent of $Z$.
\end{assumption}

\begin{assumption}\label{as:late2}
	$\Pr(T_1^* > T_0^*) = \Pr(C)>0$ and $0<\Pr(Z=1)<1$.
\end{assumption}

\begin{assumption}\label{as:late3}
	$\Pr(T_1^* < T_0^*) =\Pr(D)= 0$.
\end{assumption}

Assumption \ref{as:late1} is an exclusion restriction that requires the instrument to be unrelated to the factors affecting the outcome and/or treatment.
Intuitively, the assumption guarantees that the instrument is assigned as good as randomly.
Assumption \ref{as:late2} requires the presence of compliers.
This is a relevance condition requiring a positive effect of $Z$ on $T^*$.
Assumption \ref{as:late3} rules out the presence of defiers, which is known as a monotonicity condition in the LATE literature.

In the example of returns to schooling with four-year college proximity $Z$, the LATE is the average returns of college degrees for individuals who graduate if and only if they live close to four-year colleges.
Assumption \ref{as:late1} implies that college proximity is unrelated to the factors affecting wages with and without a college degree and potential educational attainment with and without college proximity.
Assumption \ref{as:late2} requires the presence of individuals who graduate if and only if those individuals live close to colleges.
Under Assumption \ref{as:late3}, no individuals graduate when they do not live close to a college, but they do not graduate when they do live close to a college.

Based on the identification result in \eqref{eq:LATE}, we call $\beta^*$ ``the LATE'' below for convenience of explanation, meaning that we implicitly assume that Assumptions \ref{as:late1}, \ref{as:late2}, and \ref{as:late3} hold throughout the paper.

\begin{remark}
	In practice, $Y$ or $Z$ may also contain a measurement error.
	However, a non-differential measurement error for $Y$ or $Z$ may not contaminate the LATE inference.
	See the supplementary appendix for details.
	For this reason, our analysis below presumes that $Y$ and $Z$ do not contain measurement errors.
\end{remark}

\begin{remark}
	The LATE literature without measurement errors considers the identification of a LATE parameter when $Z$ is non-binary but a general discrete variable.
	While we focus on a binary $Z$ in the main body of the paper, our proposed procedure can also be extended to such situations.
	We develop this extension in the supplementary appendix.
\end{remark}

\subsection{Identification problem due to measurement errors}\label{sec:problem}

This section explores the identification problem of the LATE $\beta^*$ in the situation where observed $T$ may be a mismeasured variable of true $T^*$.

The identification result in \eqref{eq:LATE} implicitly requires the precise measurement of true $T^*$ in addition to the identification conditions of Assumptions \ref{as:late1}, \ref{as:late2}, and \ref{as:late3}.
However, observed $T$ may contain a measurement error in practice, of which there are a few types.
For example, in the returns to schooling analysis, educational attainment may be misrecorded randomly during the process of correcting the survey data.
Such a measurement error is independent of the factors affecting the outcome.
Our proposed procedure allows this type of measurement error.
The other possibility of measurement errors is false reporting.
Individuals may misunderstand the question or have poor recall about their educational attainment when responding to a survey.
Individuals might also have an incentive to make false statements about their academic achievement to enhance their careers.
Such measurement errors may be correlated with the observables and/or unobservables.
Our analysis can allow for false reporting depending on the observables, and we can explicitly allow the measurement error to depend on the observables (see Remark \ref{remark:control} below).

To examine the misclassification problem, we define the misclassification probability:
\begin{align*}
	m_t \coloneqq \Pr(T \neq T^*| T^*=t) = \Pr(T=1-t|T^*=t) \qquad  \text{for} \quad t = 0,1.
\end{align*}
In words, $m_0$ is the probability that an individual who is actually untreated is misclassified as treated, and $m_1$ is analogous.
The misclassification probability can also be regarded as the distribution of the measurement error.

The measurement error for the treatment is non-classical in the sense that it is dependent on the true treatment.
This is because the support of the measurement error for a discrete variable depends on the true variable.
To see this, we denote the measurement error for $T$ as $U_T \coloneqq T - T^*$.
By construction, $U_T \in \{0,1\}$ given $T^*=0$ but $U_T \in \{-1,0\}$ given $T^*=1$, meaning that $U_T$ is dependent on $T^*$.
Specifically, it is easy to see that $Cov(U_T,T^*) = - (m_0 + m_1) \Pr(T^*=0) \Pr(T^*=1)$, implying that the correlation between the error and true treatment is always negative, whereas its magnitude depends on the misclassification probabilities.
We can also show that the correlation between $T^*$ and $T$ depends on the misclassification probabilities:
\begin{equation}\label{eq:cov}
\begin{split}
	Cov(T^*,T)
	= (1- m_0 - m_1) \Pr(T^*=0) \Pr(T^*=1).
\end{split}
\end{equation}
If the sum of the misclassification probabilities $m_0+m_1$ is less (greater) than one, $T$ is positively (negatively) correlated with $T^*$.

To explore the identification problem for the LATE $\beta^* = \Delta \mu/ \Delta p^*$, we examine the relationship between observable treatment probability $p_z \coloneqq E(T|Z=z) = \Pr(T=1|Z=z)$ and true treatment probability $p_z^* = E(T^*|Z=z) = \Pr(T^*=1|Z=z)$.
We have the following relationship according to the law of iterated expectations:
\begin{equation}\label{eq:pZ}
\begin{split}
	p_z
	= m_{0z} (1-p_z^*) + (1-m_{1z})p_z^*
	= m_{0z} + s_z p_z^*,
\end{split}
\end{equation}
where $m_{tz} \coloneqq \Pr(T \neq T^*|T^*=t,Z=z) = \Pr(T=1-t|T^*=t,Z=z)$ is the conditional misclassification probability and $s_z \coloneqq 1- m_{0z} - m_{1z}$ for $t,z=0,1$.
We note that $|s_z| \le 1$ by definition.
The above equation implies that
\begin{align}\label{eq:pZstar}
	p_z^* = \frac{p_z - m_{0z}}{s_z}.
\end{align}
The true treatment probability thus depends on the conditional misclassification probabilities.

The misclassification of the treatment variable brings out the serious problem that $\beta^*$ and $\Delta p^*$ cannot be point-identified based on the observables of $(Y,T,Z)$.
Equation \eqref{eq:pZ} leads to the following system:
\begin{align*}
\left\{
\begin{array}{c}
	p_0 = m_{00} + s_0 p_0^*\\
	p_1 = m_{01} + s_1 p_1^*
\end{array}
. \right.
\end{align*}
Even when $Z$ is not related to the misclassification probability, namely $m_t = m_{tz}$ for $t,z=0,1$ (this implies $s = s_0=s_1$ for $s \coloneqq 1-m_0-m_1$), we cannot identify $p_0^*$ and $p_1^*$ because there are four unknown parameters $(m_0,s,p_0^*,p_1^*)$ in the system of two equations.
As a result, neither $\beta^*$ nor $\Delta p^*$ can be identified based on $(Y,T,Z)$.

The IV estimand based on the observables may over- or underestimate $\beta^*$.
The IV estimand based on the observables is
\begin{align}\label{eq:naive}
	\beta \coloneqq \frac{\mu_1 - \mu_0}{p_1-p_0}.
\end{align}
From the relationship in \eqref{eq:pZ}, the difference between the denominators of $\beta$ and $\beta^*$ (i.e., the difference between the observable and true first-stage regressions) is expanded as
\begin{align*}
	p_1 - p_0 - (p_1^* - p_0^*)
	= (m_{00}-m_{01}) + (1+s_1)p_1^* - (1+s_0)p_0^*,
\end{align*}
which can be negative or positive depending on the misclassification probabilities.
For example, if the misclassification probabilities for the truly untreated given $Z=0$ and $Z=1$ are the same so that $m_{00}=m_{01}$, the sign of the difference between the first-stage regressions depends on that of $(m_{10}+m_{11})/(m_{00}+m_{01})-p_0^*/p_1^*$.
As a result, observable $\beta$ may over- or underestimate true $\beta^*$.

\begin{remark}
	While $\beta$ exhibits a bias for $\beta^*$ in general, it is sufficient to estimate $\beta$ if we are interested in testing the hypothesis of whether $\beta^*=0$.
	Since $\beta^*=0$ if and only if $\beta=0$, the standard IV inference based on observed $(Y,T,Z)$ allows us to examine whether the true treatment has a mean effect on the outcome for the compliers.
\end{remark}

\begin{remark}
	When the misclassification probabilities do not depend on $Z$, namely when $m_t = m_{t0} = m_{t1}$ for $t=0,1$ so that $s = s_0 = s_1$, it holds that
	\begin{equation}\label{eq:biasspecial}
	\begin{split}
	\beta
	= \frac{\mu_1 - \mu_0}{m_{01} - m_{00} + s_1 p_1^* - s_0 p_0^*}
	= \frac{\beta^*}{s}.
	\end{split}
	\end{equation}
	Since $|s| \le 1$, $|\beta|$ is an upper bound of $|\beta^*|$, but it is not the sharp bound.
	See \citet{ura2017hetero}.
\end{remark}

\section{Identification}\label{sec:identification}
This section develops the identification of the LATE with misclassification.
Section \ref{sec:identification-result} shows the identification result based on an exogenous variable $V$.
Section \ref{sec:moment} discusses the moment conditions derived from the identification result and GMM estimation.

\subsection{Identification of the LATE with misclassification}\label{sec:identification-result}
We need the following assumptions to identify the LATE and misclassification probabilities based on the use of exogenous $V$, similarly to \citet{mahajan2006identification} and \citet{lewbel2007estimation}.

\begin{assumption}[non-differential error]\label{as:nondif}
	$E(Y|T, T^*,Z,V)=E(Y|T^*,Z,V)$.
\end{assumption}

Assumption \ref{as:nondif} implies that mismeasured $T$ has no information on the mean of $Y$ once true $T^*$, $Z$, and $V$ are conditioned on.
With the measurement error $U_T = T-T^*$, this assumption means that $E(Y|U_T,T^*,Z,V)=E(Y|T^*,Z,V)$, implying that $U_T$ is a {\it non-differential} measurement error.
The assumptions of non-differential errors are popular in the misclassification literature.
The non-differential error may require that the error does not depend on the unobservables affecting outcome $Y$ and/or true treatment $T^*$.
The non-differential error also rules out a placebo effect such that the misclassified treatment affects the outcome of the individual who does not actually receive the treatment.

For example, consider the returns to schooling analysis.
Under Assumption \ref{as:nondif}, the average wages for individuals who report graduating and, indeed, not graduating are identical to that for individuals who did not graduate with precise reports.
This could be satisfied if misclassification is caused by accident or by false statements depending the observable variables.
On the contrary, the assumption might be violated if misclassification depends on the causal effect $Y_1-Y_0$ and/or unobservables affecting wages.

\begin{assumption}[monotonicity]\label{as:neq}
	$1 - m_{0z} - m_{1z} > 0$ for $z=0,1$.
\end{assumption}

Assumption \ref{as:neq} means that the sum of the misclassification probabilities is less than one.
The assumption is popular and known as the monotonicity condition in the misclassification literature.
This is satisfied when the actual reporting for the treatment has better information about the true treatment than the random reporting in which the misclassification probabilities $m_{0z}$ and $m_{1z}$ are equal to half.
Indeed, $T^*$ is positively correlated with $T$ under this assumption according to \eqref{eq:cov}, implying that the observed treatment has positive information on the true treatment.

To introduce the next assumption, we define the following shorthand notations:
\begin{align*}
	m_{tzv} &\coloneqq \Pr(T \neq T^*|T^*=t,Z=z,V=v),\\	
	p_{zv}^* &\coloneqq E(T^*|Z=z,V=v) = \Pr(T^*=1|Z=z,V=v),\\
	\tau_{zv}^* &\coloneqq E(Y|T^*=1,Z=z,V=v) - E(Y|T^*=0,Z=z,V=v),\\
	\tau_z^* &\coloneqq E(Y|T^*=1,Z=z) - E(Y|T^*=0,Z=z),
\end{align*}
for $t,z=0,1$ and $v \in supp(V) \subset \mathbb{R}$.
Here, $m_{tzv}$ is the conditional misclassification probability, $p_{zv}^*$ is the conditional true treatment probability, and $\tau_{zv}^*$ and $\tau_z^*$ are the differences in the conditional outcome means.

\begin{assumption}[exclusion restriction and relevance condition]\label{as:ex}
	For each $z=0,1$, there exists a subset $\Omega_z \subset supp(V)$ such that
	\begin{align*}
		\tau_z^* = \tau_{z v}^*, \qquad
		m_{0z} = m_{0 z v}, \qquad
		m_{1z} = m_{1 z v},
	\end{align*}
	for any $v \in \Omega_z$, and $p^*_{zv}$ depends on $v \in \Omega_z$ so that $p^*_{zv}$ is not constant in $v \in \Omega_z$.
\end{assumption}

Assumption \ref{as:ex} contains a set of exclusion restrictions requiring that exogenous $V$ does not affect $\tau_{zv}^*$ and $m_{tzv}$.
A sufficient condition of the assumption about $\tau_{zv}^*$ is that the functional form of $E(Y|T^*,Z,V)$ is given by $E(Y|T^*,Z,V) = h_1(T^*,Z) + h_2(Z,V)$ for functions $h_1$ and $h_2$.
The assumption could hold, especially when the effect of $V$ on $Y_1$ is the same as that on $Y_0$.
Remarkably, Assumption \ref{as:ex} allows a direct effect of $V$ on $Y$, so that $V$ can be a covariate.
For example, even in the linear model of $E(Y|T^*,Z,V) = \gamma_0 +  \gamma_1 T^* + \gamma_2 Z + \gamma_3 V + \gamma_4 ZV$ with coefficients $\gamma$s, the assumption is satisfied.
The other special case in which the assumption may hold is when $V$ is randomly assigned by an experiment.
Also, when $V$ is a repeated measure of the treatment, the assumption of $\tau_{zv}^*$ requires that the error for $V$ is a non-differential error.

Under Assumption \ref{as:ex}, the generating process of the misclassification also does not depend on $V$.
However, it allows the misclassification probabilities to depend on $Z$.
We note that when $V$ is a repeated measure of the treatment that may contain a measurement error, Assumption \ref{as:ex} requires that the error for $V$ is independent of the error for $T$.
Indeed, if the error for $V$ is correlated with the error for $T$, the misclassification probability $m_{tzv}$ may depend on the value of $V$.

Assumption \ref{as:ex} further includes a relevance condition under which the true treatment probability $p_{zv}^*$ depends on the value of $V$.
This holds when $V$ has a direct effect on $T^*$ or when $V$ is a repeated measure so that $V$ relates to $T^*$.
Importantly, the relevance condition is testable under the exclusion restriction that $m_{tz}=m_{tzv}$ in Assumption \ref{as:ex}.
Indeed, with the definition of the conditional treatment probability
\begin{align*}
	p_{zv} \coloneqq E(T|Z=z,V=v) = \Pr(T=1|Z=z, V=v),
\end{align*}
the same procedure for showing \eqref{eq:pZ} leads to $p_{zv} = m_{0zv} + (1-m_{0zv}-m_{1zv}) p_{zv}^*$.
This implies that $p_{zv}^* \neq p_{zv'}^*$ for $v, v' \in \Omega_z$ if and only if $p_{zv} \neq p_{zv'}$ under exclusion restriction $m_{tz}=m_{tzv}$.

To understand the practical implications of Assumption \ref{as:ex}, let us consider the returns to schooling analysis.
Suppose that $Z$ is the indicator of four-year college proximity and that $V$ is the indicator of two-year college proximity.
The exclusion restriction for $\tau_{zv}^*$ is satisfied if the mean effect of the college degree on wages for individuals' proximity to two-year colleges is identical to that for individuals who do not live close to two-year colleges.
The condition for $m_{tzv}$ holds when two-year college proximity does not determine the generation of the measurement error.
The relevance condition for $p_{zv}^*$ is satisfied when two-year college proximity is related to true educational attainment.

For the next assumption, we define the following quantity:
\begin{align*}
		\tau_{zv} \coloneqq E(Y|T=1,Z=z,V=v) - E(Y|T=0,Z=z,V=v),
\end{align*}
for $z=0,1$ and $v \in supp(V) \subset \mathbb{R}$.
Here, $\tau_{zv}$ is the difference in the conditional outcome means, which is identified from the observable data.

The following assumption includes the conditions to solve the systems of linear equations for the identification of the misclassification probabilities.
Condition (i) is for the case where $V$ takes at least three values.
On the contrary, condition (ii) allows the situation where $V$ is a binary covariate, instrument, or repeated measure of the treatment.
We remember the set $\Omega_z \subset supp(V)$ in Assumption \ref{as:ex}.

\begin{assumption}[nonsingularity]\label{as:nonsing}
	One of the following assumptions holds.
	(i) There are at least three elements $\{v_1, v_2, v_3\} \subset \Omega_z$ for each $z=0,1$ such that
	\begin{align*}
		\left( \frac{\tau_{zv_1}}{p_{zv_2}} - \frac{\tau_{zv_2}}{p_{zv_1}}\right) \left( \frac{\tau_{zv_1}}{1-p_{zv_3}} - \frac{\tau_{zv_3}}{1-p_{zv_1}} \right)
		\neq \left( \frac{\tau_{zv_1}}{1-p_{zv_2}} - \frac{\tau_{zv_2}}{1-p_{zv_1}} \right) \left( \frac{\tau_{zv_1}}{p_{zv_3}} - \frac{\tau_{zv_3}}{p_{zv_1}} \right).
	\end{align*}
	(ii) It holds that $m_t=m_{t0} = m_{t1}$ for each $t=0,1$ and there are at least two elements $\{v_1, v_2\} \subset \Omega_0 \cap \Omega_1$ such that
	\begin{align*}
		\left( \frac{\tau_{0v_1}}{p_{0v_2}} - \frac{\tau_{0v_2}}{p_{0v_1}}\right) \left( \frac{\tau_{1v_1}}{1-p_{1v_2}} - \frac{\tau_{1v_2}}{1-p_{1v_1}} \right) \neq \left( \frac{\tau_{0v_1}}{1-p_{0v_2}} - \frac{\tau_{0v_2}}{1-p_{0v_1}} \right) \left( \frac{\tau_{1v_1}}{p_{1v_2}} - \frac{\tau_{1v_2}}{p_{1v_1}} \right).
	\end{align*}
\end{assumption}

Assumption \ref{as:nonsing} contains somewhat technical inequalities that ensure the unique solutions for the systems of linear equations.
Note that the inequalities are testable in principle since their components are identified by the data.
Also, as shown in the supplementary appendix, the necessary and sufficient conditions of the inequalities are $\tau_z^* \neq 0$ and $m_{0z} + m_{1z} \neq 1$.
For example, in the returns to schooling example, the inequalities hold if and only if average wages for college graduates are different from those for individuals without college degrees.

Condition (ii) also requires the additional exclusion restriction that the misclassification probabilities do not depend on $Z$.
We need the additional exclusion restriction since binary $V$ is less informative for the distribution of $T^*$ than non-binary $V$ under the exclusion restrictions in Assumption \ref{as:ex}.
We note that \citet{calvi2017women}, \citet{ditraglia2015mis}, and \citet{ura2017hetero} also assume the same exclusion restriction.

The following theorem states the main identification result of this study.

\begin{theorem}\label{thm:iden}
	Suppose that Assumptions \ref{as:late1}, \ref{as:late2}, \ref{as:late3}, \ref{as:nondif}, \ref{as:neq}, \ref{as:ex}, and \ref{as:nonsing} are satisfied.
	The LATE $\beta^*$, true treatment probability $p_z^*$, and misclassification probabilities $m_{0z}$ and $m_{1z}$ for each $z = 0,1$ are identified.
\end{theorem}

The main idea behind the identification is, similar to \citet{mahajan2006identification} and \citet{lewbel2007estimation}, to construct moment equalities whose number is no fewer than the unknown parameters.
The idea can be understood by a simple sketch of the proof of Theorem \ref{thm:iden}.
This proof depends on whether we assume Assumption \ref{as:nonsing} (i) or (ii).
Under Assumptions \ref{as:nondif}, \ref{as:neq}, \ref{as:ex}, and \ref{as:nonsing} (i), we can show the following system of equations for each $z=0,1$:
\begin{align}\label{eq:sys1}
	\left\{
	\begin{array}{c}
	B_{0z} w_{0zv_1v_2} + B_{1z} w_{1zv_1v_2} + w_{2zv_1v_2} = 0\\
	B_{0z} w_{0zv_1v_3} + B_{1z} w_{1zv_1v_3} + w_{2zv_1v_3} = 0
	\end{array}
	\right.,
\end{align}
where the $B$s are parameters to be identified in our analysis and $w$s are identified by data.
The definitions of the $B$s and $w$s are given in \eqref{eq:Bw} in the proof of Theorem \ref{thm:iden}.
Since the $w$s are identified from the observable data and Assumption \ref{as:nonsing} (i) guarantees the existence of the unique solutions of the $B$s, the $B$s are identified.
Similarly, under the assumption of $m_{t}=m_{t0} = m_{t1}$ for $t=0,1$ in Assumption \ref{as:nonsing} (ii), we instead have
\begin{align}\label{eq:sys2}
		\left\{
		\begin{array}{c}
		B_0 w_{00v_1v_2} + B_1 w_{10v_1v_2} + w_{20v_1v_2} = 0\\
		B_0 w_{01v_1v_2} + B_1 w_{11v_1v_2} + w_{21v_1v_2} = 0
		\end{array}
		\right.,
\end{align}
We note that the $B$s in \eqref{eq:sys2} do not depend on $z$ unlike the $B$s in \eqref{eq:sys1}.
In this case, Assumption \ref{as:nonsing} (ii) guarantees that the $B$s are identified as the solution of simultaneous equations \eqref{eq:sys2}.
As a next step, the identification of the $B$s leads to identifying the misclassification probabilities.
Specifically, under Assumption \ref{as:nonsing} (i), we have
\begin{align*}
	s_z = \sqrt{(B_{0z} - B_{1z} + 1)^2 - 4B_{0z}},
	\qquad
	m_{0z}= (B_{0z}-B_{1z}+1-s_{z})/2,
	\qquad m_{1z} = 1- m_{0z} - s_z,
\end{align*}
for each $z=0,1$.
On the contrary, under Assumption \ref{as:nonsing} (ii), we have
\begin{align*}
	s = \sqrt{(B_0 - B_1 + 1)^2 - 4B_0},
	\qquad
	m_0= (B_0-B_1+1-s)/2,
	\qquad
	m_1 = 1- m_0 - s.
\end{align*}
These equations mean that the misclassification probabilities are identified.
Finally, the true first-stage regression is identified based on the information on the misclassification probability and observable treatment probability.
Under Assumption \ref{as:nonsing} (i), we have
\begin{align*}
	\Delta p^* = \frac{p_1 - m_{01}}{1-m_{01} - m_{11}} - \frac{p_0 - m_{00}}{1-m_{00} - m_{10}},
\end{align*}
and, under Assumption \ref{as:nonsing} (ii), we have
\begin{align*}
	\Delta p^* = \frac{p_1 - p_0}{1-m_0-m_1}.
\end{align*}
Hence, the LATE $\beta^* = \Delta \mu/\Delta p^*$ is identified.

\begin{remark}\label{remark:control}
	The proposed analysis can be extended to situations in which other control variables are observed.
	Suppose we observe a vector of the control variables $S \in \mathbb{R}^d$ in addition to $(Y,T,Z,V)$.
	Let the parameter of interest be the LATE conditional on the control variables:
	$E(Y_1 - Y_0|C, S=s) = E(Y_1 - Y_0|T_1^* > T_0^*, S=s)$ where $s \in supp(S)$.
	If Assumptions \ref{as:late1}, \ref{as:late2}, and \ref{as:late3} are satisfied conditional on $S$, it holds that
	\begin{align*}
		E(Y_1 - Y_0|T_1^* > T_0^*, S=s) = \frac{E(Y|Z=1, S=s) - E(Y|Z=0,S=s)}{E(T^*|Z=1, S=s) - E(T^*|Z=0,S=s)}.
	\end{align*}
	The right-hand side is the IV estimand conditional on $S=s$, which is identified if $(Y,T^*,Z,S)$ is observed without a measurement error.
	In addition, if Assumptions \ref{as:nondif}, \ref{as:neq}, \ref{as:ex}, and \ref{as:nonsing} hold conditional on $S=s$, the identification result in Theorem \ref{thm:iden} holds conditional on $S=s$.
	Importantly, owing to the presence of control variables $S$, we can explicitly allow the generation of the measurement error to depend on the observables.
	For example, Assumption \ref{as:nondif} may be replaced with the non-differential error conditional on $S$:
	$E(Y|T, T^*,Z,V,S) = E(Y|T^*,Z,V,S)$.
	This implies that the measurement error for the treatment does not affect the mean of $Y$ once $T^*$ and $S$ are conditioned on.
\end{remark}

\begin{remark}
	When $V$ is a binary instrument as well as $Z$, we can consider the analysis in which the roles of $Z$ and $V$ are changed.
	However, we should be careful in interpreting empirical results in this situation since the identified parameters depend on the roles of $Z$ and $V$.
	We discuss this issue in detail in the supplementary appendix.
\end{remark}

\begin{remark}\label{remark:covariance}
	The exclusion restriction and relevance condition in Assumption \ref{as:ex} relate to conditional covariance restrictions in \citet{caetano2017identifying}.\footnote{
	The author appreciates an editor's comment on the relevance discussed here.}
	We here discuss the relationship between our restriction for $\tau_{zv}^*$ and their covariance restriction.
	Suppose that $E(Y|T^*,Z,V) = h_1(T^*,Z) + h_2(Z,V)$ that is the sufficient condition of our restriction as discussed above.
	Let $R^* \coloneqq (T^*, Z)^\top$ and $W \coloneqq (Z, V)^\top$.
	For any function $q$, we have
	\begin{align*}
		Cov(Y,  q(R^*, W)| W) = Cov( h_1(R^*), q(R^*, W)| W),
	\end{align*}
	which is a type of conditional covariance restrictions in \citet{caetano2017identifying}.
	Nonetheless, it should be noted that in our setting $R^*$ is partially unobserved due to misclassification, while they study models without measurement errors.
\end{remark}

\subsection{Moment conditions}\label{sec:moment}
We derive the moment conditions based on the identification result in Theorem \ref{thm:iden}.
These moment conditions lead to estimation procedures such as the non-linear GMM estimation and empirical likelihood estimation.
Here, we focus on the moment conditions under Assumption \ref{as:nonsing} (i) since those under Assumption \ref{as:nonsing} (ii) are simpler.
We assume that $V$ is a discrete random variable since covariates, instruments, and repeated measures are often discrete in practice.
We can easily derive similar moment conditions when $V$ is continuously distributed and/or when other control variables exist.

Define the vector of the observable variables $X \coloneqq (Y,T,Z,V)^\top$.
For simplicity, suppose that discrete $V$ takes $K$ values in $supp(V) = \Omega_0 = \Omega_1 = \{v_1, v_2, \dots, v_K\}$, where $\Omega_0$ and $\Omega_1$ are introduced in Assumptions \ref{as:ex} and \ref{as:nonsing}.
Define a vector of the $K+3$ parameters:
\begin{align*}
	\theta_0^{(z)} \coloneqq (m_{0z}, m_{1z},p^*_{zv_1},p^*_{zv_2},\dots,p^*_{zv_K},\tau_z^*)^\top.
\end{align*}
The vector $\theta_0^{(z)}$ for $z=0,1$ contains the parameters conditional on $Z=z$ introduced in Section \ref{sec:identification-result}.
The vector of the $2K+9$ parameters to be estimated is
\begin{align}\label{eq:q0}
	\theta_0 \coloneqq (\beta^*, \Delta p^*, r,\theta_0^{(0)\top}, \theta_0^{(1)\top})^\top,
\end{align}
where $r \coloneqq E(Z)$.
The parameters except for the LATE $\beta^*$ and the true first-stage regression $\Delta p^*$ are nuisance parameters to overcome the misclassification problem.
Nonetheless, as well as $\beta^*$ and $\Delta p^*$, the misclassification probability $m_{tz}$ could be of interest in practice.
Let $\theta$ be a general parameter value in the parameter space $\Theta \subset \mathbb{R}^{2K+9}$ to which $\theta_0$ also belongs.

Let $g(X, \theta_0)$ be the vector valued function with $4K+3$ elements, which are the following components:
for $k=1,2,\dots,K$ and $z=0,1$,
\begin{equation}\label{eq:moment}
\begin{split}
	& r - Z,\\
	& \Big( m_{0z} + (1-m_{0z}-m_{1z})p_{zv_k}^* - T \Big) I_{zv_k},\\
	& \left( \tau_z^* + \frac{YT-(1-m_{1z})p^*_{z v_k}\tau_z^*}{m_{0z} + (1-m_{0z}-m_{1z})p^*_{zv_k}} - \frac{Y(1-T)+(1-m_{0z})(1-p^*_{zv_k})\tau_z^*}{1-(m_{0z}+(1-m_{0z}-m_{1z})p^*_{z v_k})}\right) I_{zv_k},\\
	& \Delta p^* - \left( \frac{TZr^{-1} -m_{01}}{1-m_{01}-m_{11}} - \frac{T(1-Z)(1-r)^{-1} -m_{00}}{1-m_{00}-m_{10}} \right),\\
	& \beta^* - \frac{YZr^{-1} - Y(1-Z)(1-r)^{-1}}{\Delta p^*},
\end{split}
\end{equation}
where $I_{zv_k} \coloneqq \mathbf{1}(Z=z,V=v_k)$ is the indicator.

The following theorem shows that the moment condition is $E[g(X,\theta_0)]=0$ with the unique solution $\theta_0 \in \Theta$.
This is directly shown by the identification result in Theorem \ref{thm:iden}.

\begin{theorem}\label{thm:gmm}
	Suppose that Assumptions \ref{as:late1}, \ref{as:late2}, \ref{as:late3}, \ref{as:nondif}, \ref{as:neq}, \ref{as:ex}, and \ref{as:nonsing} (i) hold with $\Omega_0 = \Omega_1 = \{v_1, v_2, \dots, v_K\}$.
	Then, it holds that $E[g(X, \theta_0)]=0$ and $\theta_0$ is its unique solution in the sense that $E[g(X,\theta)] \neq 0$ for any $\theta \in \Theta$ such that $\theta \neq \theta_0$.
\end{theorem}

The number of overidentification restrictions depends on the number of elements in support of $V$, that is, $K$.
As stated above, the numbers of parameters and moment equations are $2K+9$ and $4K+3$, respectively, under Assumption \ref{as:nonsing} (i).
Hence, there are $2K-6$ overidentification restrictions and $\theta_0$ is just-identified when $K=3$.

Similarly, under Assumption \ref{as:nonsing} (ii), we can show that there are $2K-4$ overidentification restrictions, and just-identification is achieved when $K=2$.

\begin{remark}
	With a random sample $\{X_i\}_{i=1}^n$ of $X$, we can estimate $\theta_0$ via the non-linear GMM estimation based on $E[g(X,\theta_0)]=0$.
	The GMM estimator is $\sqrt{n}$-consistent and asymptotically normal under regularity conditions.
	We can implement confidence interval estimation and hypothesis testing for $\theta_0$ in the usual manner.
	We can also achieve the optimal GMM estimation based on the optimal weighting matrix as usual.
	It is important to note that the overidentification test allow us to examine the validity of our identification conditions when there are overidentification restrictions.
	See, for example, \citet[Chapter 14]{wooldridge2010econometric} for detailed procedures.
\end{remark}

\section{Monte Carlo simulations}\label{sec:monte}
This section reports the results of the Monte Carlo simulations.
The simulations are conducted by {\ttfamily R} with 2,000 simulation replications.

\paragraph{DGP.}
By using sample sizes of 200, 500, and 1,000, we generate the random variables by adopting the following six data-generating processes.
For all designs, we generate $Z \in \{0,1\}$ with probabilities $\Pr(Z=0) = \Pr(Z=1) = 0.5$ and the following unobservables affecting the outcome and the true treatment:
\begin{align*}
	\left(
	\begin{array}{c}
	U_1 \\
	U_2
	\end{array}
	\right)
	\sim
	N \left(
	\left(
	\begin{array}{c}
	0 \\
	0
	\end{array}
	\right),
	\left(
	\begin{array}{cc}
	1 & 0.05 \\
	0.05 & 0.5
	\end{array}
	\right)
	\right).
\end{align*}

In designs 1 and 2, we presume $V$ to be a covariate and consider two different designs for $Y$.
Assuming Assumption \ref{as:nonsing} (ii), we generate binary $V$ with probabilities $\Pr(V=0)=\Pr(V=1)=0.5$.
The true treatment is generated by the probit model $T^*=\mathbf{1}(-1 + Z + V - U_1>0)$.
The observed treatment $T$ is misclassified independently of the other variables with misclassification probability $m_t = \Pr(T \neq T^*|T^*=t)=0.25$ for each $t=0,1$.
Two designs are considered for the outcome variable:
\begin{align*}
	\text{Design 1:} \quad Y = 1+T^* + 0.3 V + U_2,
	\qquad
	\text{Design 2:} \quad Y = U_2(2 U_2 - 1)T^* + 0.3 V + U_2.
\end{align*}
Design 1 considers the homogeneous causal effect, whereas the causal effect in design 2 is heterogeneous depending on unobserved $U_2$.
In both designs, $V$ directly affects $Y$.

In designs 3 and 4, we presume $V$ to be an instrument.
Generating binary $V$ with probabilities $\Pr(V=0)=\Pr(V=1)=0.5$, the true treatment is generated by $T^*=\mathbf{1}(-1 + Z + V - U_1>0)$.
The misclassification probability is $m_t = \Pr(T \neq T^*|T^*=t)=0.25$ for each $t=0,1$.
The outcome is generated by
\begin{align*}
	\text{Design 3:} \quad Y = 1+T^*+ U_2,
	\qquad
	\text{Design 4:} \quad Y = U_2(2 U_2 - 1)T^*+ U_2.
\end{align*}

In designs 5 and 6, we presume $V$ to be a binary repeated measure.
With the true treatment $T^*=\mathbf{1}(-0.5 + Z - U_1 > 0)$, the observed treatment $T$ and repeated measure $V$ have the misclassification probabilities $m_t = \Pr(T \neq T^* | T^* = t) = 0.25$ and $\Pr(V \neq T^* | T^* = t) = 0.3$ for $t= 0, 1$ independently of other variables.
Note that $V \neq T$ in general.
The outcome is generated by
\begin{align*}
	\text{Design 5:} \quad Y=1+T^* + U_2,
	\qquad
	\text{Design 6:} \quad Y = U_2(2 U_2 - 1)T^* + U_2.
\end{align*}

\paragraph{Estimators.}
We consider two estimators.
The first is the GMM estimator based on the identification result developed in this study.
The second is the naive IV estimator $\beta$ in \eqref{eq:naive}, which is a benchmark estimator.

In this simulation, the GMM estimation involves the 11 parameters to be estimated, that is, $\theta_0 = (\beta^*, \Delta p^*, r,m_0,p_{00}^*,p_{01}^*,\tau_0^*,m_1,p_{10}^*,p_{11}^*,\tau_1^*)^\top$.
We focus on the parameters of interest of the LATE $\beta^*$, first-stage regression $\Delta p^*$, and misclassification probabilities $m_0$ and $m_1$.
The moment condition $E[g(X, \theta_0)]=0$ is composed of 11 elements.
Since $\theta_0$ is just-identified here, we select the identity matrix as the weighting matrix.

\paragraph{Result.}
Tables \ref{table:monte:LATE}, \ref{table:monte:first}, and \ref{table:monte:m0} summarize the Monte Carlo simulation results for $\beta^*$, $\Delta p^*$, $m_0$, and $m_1$.
Since the main parameter of interest is $\beta^*$, we focus on the results reported in Table \ref{table:monte:LATE}.
The table reports the true value and biases, standard deviations (SDs), and root mean squared errors (RMSEs) for the proposed GMM estimator and naive IV estimator ignoring the measurement error.
It also indicates the coverage probabilities (CPs) of the 95\% confidence intervals based on the asymptotic normal approximations for the estimators with heteroscedasticity robust standard errors.

The performance of our GMM estimation is successful.
The bias and SD of the GMM estimator are satisfactory in all designs.
For example, the biases of the GMM estimator for the LATE when $n=1000$ can be about 10\% of the true values.
The RMSE of the GMM estimator in each design is also moderate and becomes smaller as the sample size increases, which is expected owing to the asymptotics of the GMM estimator.
Further, the CPs of the 95\% confidence interval are close to 0.95 even with small sample sizes.

Table \ref{table:monte:LATE} also shows that the naive IV estimator $\beta$ based on the observables exhibits large biases in all designs.
In each design, the bias is over 100\% of the true value of the LATE, and the naive IV estimator overestimates the LATE.
The result is driven by the identification failure of the naive IV estimand for the LATE, as we discuss in Section \ref{sec:problem}, and the magnitude of the bias is consistent with our theoretical investigation in \eqref{eq:biasspecial}.

\section{Empirical illustration}\label{sec:empirical}
We illustrate our proposed procedure by analyzing returns to schooling.
We use the same dataset used by \citet{card1993using}.
The data are originally drawn from the National Longitudinal Survey of Young Men, and the number of the observations is 3,010.
See \citet{card1993using} for the details on the features of the dataset.

The variables we use are log hourly wages ($lwage$), the schooling indicator ($college$) whether years of schooling are no less than 14, that is, the indicator corresponding degrees greater than two-year college degrees, the indicator of two-year college proximity ($proximity2$), and the indicator of four-year college proximity ($proximity4$).
We call individuals with $college = 1$ college graduates, while we recognize that those individuals might include college dropouts.
We may also use the other control variables contained in the original dataset such as age, although we do not use those variables here for two reasons.
First, to incorporate those variables into the GMM estimation, we need functional form specifications such as linear models for the functions in the parameters, which might cause misspecification problems.
Second, the results of the naive estimation ignoring the measurement error are not sensitive to whether we control for those variables.

Table \ref{table:empirical:descriptive} presents the descriptive statistics for our variables.
Table \ref{table:empirical:ign} reports the results of the naive OLS and IV estimation ignoring the presence of the measurement error.

For our proposed inference, we utilize two indicators of college proximity to deal with the endogeneity and misclassification problems simultaneously.
Our main result in this illustration uses $proximity4$ as $Z$ and $proximity2$ as $V$.
This specification is based on the fact that the observed first-stage regression when using $proximity4$ as $Z$ is significantly stronger than that when using $proximity2$ as $Z$, as shown in columns (4) and (5) in Table \ref{table:empirical:ign}.
Nonetheless, we also present the GMM estimates when using $proximity2$ as $Z$ and $proximity4$ as $V$ in Table \ref{table:empirical:gmm2} as an auxiliary result; however, these estimates are less precise than our main estimates below.

The validity of the exclusion restrictions and relevance condition in Assumption \ref{as:ex} should be examined since they are especially important among our identification conditions.
These exclusion restrictions require that the effect of the true college degree on wages and the distribution for misreporting the college degree do not depend on whether individuals live close to two-year colleges.
Importantly, college proximity can affect wages for our inference, which would be desirable since several studies point out that college proximity might be related to factors affecting wages (e.g., \citealp{carneiro2002evidence}).
The relevance condition requires that the true college degree is related to two-year college proximity.
Its validity is statistically testable as discussed in Section \ref{sec:identification}; columns (6) and (7) in Table \ref{table:empirical:ign} present the results of such a test.
Column (6) shows evidence of $p_{11}^* \neq p_{10}^*$ at a significance level of 5\%.
Column (7) shows evidence of $p_{01}^* \neq p_{00}^*$ at a significance level of 10\%, owing to the small sample size of the data with $proximity4 = 0$, which leads to a relatively large standard error.
The estimates in columns (6) and (7) are non-negligible amounts, meaning that they imply the validity of our relevance condition.

Table \ref{table:empirical:gmm1} reports the result of our GMM estimation when using $proximity4$ as $Z$ and $proximity2$ as $V$.
The GMM estimate for the LATE shows that having a college degree increases average wages for compliers by about 42\%, whereas the naive estimation shows an average effect of about 131\%.
This severe difference is caused by the misclassification problem, namely that the observed first-stage regression ignoring the measurement error underestimates the true first-stage regression.
These results demonstrate the importance of taking account of the misclassification problem as well as the usefulness of our GMM inference.

Table \ref{table:empirical:gmm1} also reports the estimates for the misclassification probabilities, showing an interesting result that individuals with a true college degree tend to misreport more than those without.
These estimated misclassification probabilities might seem to be significantly high, but they could be reliable estimates since the estimated effect of being a college graduate from the GMM estimation is more consistent with existing empirical results than is the naive IV estimate.
For example, by using other datasets, \citet{kane1999estimating}, \citet{lewbel2007estimation}, and \citet{battistin2014misreported} show that the average effects of undergraduate education on wages are about 25\%, 37\%, and 27\%, respectively if taking account of the presence of misreported schooling.

\section{Conclusion}\label{sec:conclusion}
This study presents novel point-identification for the LATE when the binary endogenous treatment may contain a measurement error.
Since the measurement error must be non-classical by construction, the standard IV estimator cannot consistently estimate the LATE.
To correct the bias due to the measurement error, we build on the results presented by \citet{mahajan2006identification} and \citet{lewbel2007estimation} to point-identify the LATE, true first-stage regression, and misclassification probabilities based on the use of an exogenous variable such as a covariate, instrument, or repeated measure of the treatment.
The moment conditions derived from the identification lead to the GMM estimator for the parameters with asymptotically valid inferences.
The simulations and empirical illustration demonstrate the usefulness of the proposed inference.

Several important future research topics can be proposed on the basis of our findings.
First, it is desirable to develop point-identification for the LATE with a {\it differential} (i.e., endogenous) measurement error for the treatment.
The measurement error is differential when it is related to the outcome variable, even conditional on the true variable.
How to handle such an error would be of interest but challenging.
Second, it would be of interest to develop inferences for the LATE with a measurement error when the treatment is a general discrete variable.
Without a measurement error for the discrete treatment, the IV estimand identifies the LATE with the variable treatment intensity (\citealp{angrist1995two}).
Our inference may be extended to such situations.
Third, the proposed identification may be extended to other settings such as regression discontinuity (RD) designs with the mismeasured treatment.
Since RD designs require local inferences around thresholds, we cannot directly use our proposed estimation for RD inferences.
The author develops this topic as another project (\citealp{yanagi2015regression}).

\appendix

\section{Appendix: Proofs of the theorems}\label{sec:proof}
This appendix contains the proofs of Theorems \ref{thm:iden} and \ref{thm:gmm}.
In the following, we write $W = (Z,V)^\top$ for notational convenience.

\subsection{Proof of Theorem \ref{thm:iden}}\label{sec:proof1}
The outline of the proof is an extension of the proof for Theorems 1 and 2 in \citet{lewbel2007estimation}.
First, we examine the relationship between identified parameter $\tau_W = E(Y|T=1,W) - E(Y|T=0,W)$ and true version $\tau_W^* =  E(Y|T^*=1,W) - E(Y|T^*=0,W)$ under Assumption \ref{as:nondif}.
Second, we clarify that the relationship between $\tau_W$ and $\tau_W^*$ involves misclassification probabilities $m_{0z}$, $m_{1z}$, and $s_z =  1-m_{0z}-m_{1z}$ under Assumptions \ref{as:nondif}, \ref{as:neq}, and \ref{as:ex}.
Third, we show that $m_{0z}$, $m_{1z}$, and $s_z$ are identified under Assumption \ref{as:nonsing} (i) or (ii).
Finally, we argue that the LATE $\beta^*$ is identified based on identified parameters $m_{0z}$, $m_{1z}$, and $s_z$.

\paragraph{Step 1.}
Assumption \ref{as:nondif} means that
\begin{equation*}
\begin{split}
	E(Y|T, T^*,W)
	&= E(Y|T^*,W) \\
	&= T^* E(Y|T^*=1,W) + (1-T^*) E(Y|T^*=0,W)\\
	&= E(Y|T^*=0,W) +T^* \cdot \tau_W^*.
\end{split}
\end{equation*}
The law of iterated expectations leads to $E(Y|T,W) = E(Y|T^*=0,W) + E(T^*|T,W) \tau_W^*$ so that we have
\begin{equation}\label{eq:tau}
\begin{split}
	\tau_W
	&= E(Y|T=1,W) - E(Y|T=0,W) \\
	&= [E(T^*|T=1,W)- E(T^*|T=0,W)  ]\tau_W^*.
\end{split}
\end{equation}

\paragraph{Step 2.}
We next examine $E(T^*|T=t,W) = \Pr(T^*=1|T=t,W)$ for $t=0,1$:
\begin{align*}
	\Pr(T^*=1|T=t,W) = \frac{\Pr(T=t|T^*=1,W)\Pr(T^*=1|W)}{\Pr(T=t|W)}.
\end{align*}
Thus, we can write
\begin{align}\label{eq:Estar}
	E(T^*|T=0,W)= \frac{m_{1W} \cdot p_W^*}{1-p_W}, \qquad
	E(T^*|T=1,W)=\frac{(1-m_{1W})p_W^*}{p_W}.
\end{align}
Further, we have
\begin{equation}\label{eq:pW}
\begin{split}
	p_W = \Pr(T=1|W)
	&= m_{0W} (1-p_W^*) + (1-m_{1W})p_W^*\\
	&= m_{0W} + ( 1-m_{0W} - m_{1W} )p_W^*.
\end{split}
\end{equation}
This leads to
\begin{align}\label{eq:pstar}
	p_W^* = \frac{p_W - m_{0W}}{1 - m_{0W} - m_{1W}},
\end{align}
under the assumption that $1 - m_{0W} - m_{1W} \neq 0$ (Assumption \ref{as:neq}).

By substituting \eqref{eq:Estar} and \eqref{eq:pstar} into \eqref{eq:tau} and rearranging the equation, we get
\begin{align}\label{eq:M}
	\tau_W = M(m_{0W},m_{1W},p_W ) \tau_W^*,
\end{align}
where we define
\begin{equation*}
	M(m_{0W},m_{1W},p_W) \coloneqq \frac{1}{1-m_{0W} - m_{1W}} \left(1-\frac{m_{0W} (1-m_{1W})}{p_W} - \frac{(1-m_{0W})m_{1W}}{1-p_W} \right).
\end{equation*}

\paragraph{Step 3.}
Under Assumption \ref{as:ex}, \eqref{eq:M} implies that for each $z \in \{0,1\}$ and every $v, v' \in \Omega_z$
\begin{align*}
	\tau_{zv} = M(m_{0z},m_{1z},p_{zv}) \tau_z^*,
	\qquad
	\tau_{zv'} = M(m_{0z},m_{1z},p_{zv'}) \tau_z^*.
\end{align*}
This implies that $\tau_{zv} M(m_{0z},m_{1z},p_{zv'}) - \tau_{zv'} M(m_{0z},m_{1z},p_{zv}) = 0$ under Assumption \ref{as:neq}.
This equation can be rearranged as follows:
\begin{align*}
& \tau_{zv} \left(1-\frac{m_{0z} (1-m_{1z})}{p_{zv'}} - \frac{(1-m_{0z})m_{1z}}{1-p_{zv'}} \right) - \tau_{zv'} \left(1-\frac{m_{0z} (1-m_{1z})}{p_{zv}} - \frac{(1-m_{0z})m_{1z}}{1-p_{zv}} \right) = 0 \\
& \Longleftrightarrow m_{0z}(1-m_{1z}) \left(\frac{\tau_{zv}}{p_{zv'}} - \frac{\tau_{zv'}}{p_{zv}}\right) + (1-m_{0z}) m_{1z}  \left(\frac{\tau_{zv}}{1-p_{zv'}} - \frac{\tau_{zv'}}{1-p_{zv}}\right) + (\tau_{zv'} - \tau_{zv}) = 0\\
& \Longleftrightarrow B_{0z} w_{0zvv'} + B_{1z} w_{1zvv'} + w_{2zvv'} = 0,
\end{align*}
where we define
\begin{equation}\label{eq:Bw}
\begin{split}
	& B_{0z} \coloneqq m_{0z}(1-m_{1z}), \qquad B_{1z} \coloneqq (1-m_{0z})m_{1z},\\
	& w_{0zvv'} \coloneqq \frac{\tau_{zv}}{p_{zv'}} - \frac{\tau_{zv'}}{p_{zv}}, \qquad
	w_{1zvv'} \coloneqq \frac{\tau_{zv}}{1-p_{zv'}} - \frac{\tau_{zv'}}{1-p_{zv}},\qquad
	w_{2zvv'} \coloneqq \tau_{zv'} - \tau_{zv}.
\end{split}
\end{equation}
Note that the $w$s are identified by the observables of $(Y,T,Z,V)$.

The remaining proof depends on whether we assume Assumption \ref{as:nonsing} (i) or (ii).

\paragraph{Step 4 under Assumption \ref{as:nonsing} (i).}
We show the identification of $B_{0z}$ and $B_{1z}$ under Assumption \ref{as:nonsing} (i).
Then, $\Omega_z$ contains at least three elements $(v_1,v_2,v_3)$ and we have the system of two linear equations:
\begin{align*}
	\left\{
	\begin{array}{c}
		B_{0z} w_{0zv_1v_2} + B_{1z} w_{1zv_1v_2} + w_{2zv_1v_2} = 0\\
		B_{0z} w_{0zv_1v_3} + B_{1z} w_{1zv_1v_3} + w_{2zv_1v_3} = 0
	\end{array}
	\right..
\end{align*}
The system can be uniquely solved for unknown parameter $(B_{0z},B_{1z})$ as long as matrix $((w_{0zv_1v_2}, w_{0zv_1v_3})^\top, (w_{1zv_1v_2}, w_{1zv_1v_3})^\top)$ is non-singular.
The necessary and sufficient condition of the non-singularity is the non-zero determinant of the matrix, i.e.,
\begin{align*}
	\left( \frac{\tau_{zv_1}}{p_{zv_2}} - \frac{\tau_{zv_2}}{p_{zv_1}}\right) \left( \frac{\tau_{zv_1}}{1-p_{zv_3}} - \frac{\tau_{zv_3}}{1-p_{zv_1}} \right) \neq \left( \frac{\tau_{zv_1}}{1-p_{zv_2}} - \frac{\tau_{zv_2}}{1-p_{zv_1}} \right) \left( \frac{\tau_{zv_1}}{p_{zv_3}} - \frac{\tau_{zv_3}}{p_{zv_1}} \right).
\end{align*}
Hence, $(B_{0z},B_{1z})$ is identified under Assumption \ref{as:nonsing} (i).

We next show the identification of $s_z = 1-m_{0z}-m_{1z}$.
The equation $B_{tz}=m_{tz}(1-m_{1-t,z})$ in \eqref{eq:Bw} for $t=0,1$ implies that $(s_z+m_{0z})m_{0z}=B_{0z}$ and $2 m_{0z} = B_{0z} - B_{1z} + 1- s_z$.
Substituting the second into the first provides
\begin{equation*}
\begin{split}
	&\left(s_z+\frac{B_{0z} - B_{1z} + 1- s_z}{2}\right) \frac{B_{0z} - B_{1z} + 1- s_z}{2} =B_{0z}\\
	& \Longleftrightarrow s_z=  \sqrt{(B_{0z} - B_{1z} + 1)^2 - 4B_{0z}},
\end{split}
\end{equation*}
under Assumptions \ref{as:neq} and \ref{as:ex}.
Hence, $s_z$ is identified by the identification of the $B$s.
Since $s_z$, $B_{0z}$, and $B_{1z}$ are identified, $m_{0z}$ and $m_{1z}$ are also identified by $m_{0z}= (B_{0z}-B_{1z}+1-s_{z})/2$ and $s_{z}=1-m_{0z}-m_{1z}$.

Finally, we argue that the LATE $\beta^*$ is identified based on the above steps.
Since it holds that $p_z= m_{0z} + (1-m_{0z} - m_{1z})p_z^*$,
the true treatment probability and true first-stage regression are identified based on
\begin{align}\label{eq:delta_ps}
	p_z^* = \frac{p_z - m_{0z}}{1-m_{0z} - m_{1z}}, \qquad
	p_1^* - p_0^* = \frac{p_1 - m_{01}}{1-m_{01} - m_{11}} - \frac{p_0 - m_{00}}{1-m_{00} - m_{10}},
\end{align}
by identified parameters $p_z$, $m_{0z}$, and $m_{1z}$.
Hence, we have shown that
\begin{align}\label{eq:taustar}
	\beta^* = \frac{\mu_1-\mu_0}{p_1^*-p_0^*},
\end{align}
is identified since the numerator is identified by the data.

\paragraph{Step 4 under Assumption \ref{as:nonsing} (ii).}
We show the identification of $B_0 \coloneqq m_0(1-m_1)$ and $B_1 \coloneqq m_1(1-m_0)$ under Assumption \ref{as:nonsing} (ii).
With the exclusion restriction of $m_{tz}=m_{t}$ in Assumption \ref{as:nonsing} (ii), we have the system of two linear equations:
\begin{align*}
	\left\{
	\begin{array}{c}
		B_0 w_{00v_1v_2} + B_1 w_{10v_1v_2} + w_{20v_1v_2} = 0\\
		B_0 w_{01v_1v_2} + B_1 w_{11v_1v_2} + w_{21v_1v_2} = 0
	\end{array}
	\right..
\end{align*}
The system can be uniquely solved for unknown parameter $(B_0,B_1)$ as long as matrix $((w_{00v_1v_2}, w_{01v_1v_2})^\top, (w_{10v_1v_2}, w_{11v_1v_2})^\top)$ is non-singular.
The necessary and sufficient condition of the non-singularity is the non-zero determinant of the matrix, i.e.,
\begin{align*}
	\left( \frac{\tau_{0v_1}}{p_{0v_2}} - \frac{\tau_{0v_2}}{p_{0v_1}}\right) \left( \frac{\tau_{1v_1}}{1-p_{1v_2}} - \frac{\tau_{1v_2}}{1-p_{1v_1}} \right) \neq \left( \frac{\tau_{0v_1}}{1-p_{0v_2}} - \frac{\tau_{0v_2}}{1-p_{0v_1}} \right) \left( \frac{\tau_{1v_1}}{p_{1v_2}} - \frac{\tau_{1v_2}}{p_{1v_1}} \right).
\end{align*}
Hence, $(B_0,B_1)$ is identified under Assumption \ref{as:nonsing} (ii).

We also show the identification of $\beta^*$.
$B_t=m_t(1-m_{1-t})$ for $t=0,1$ implies that $(s+m_0)m_0=B_0$ and $2 m_0 = B_0 - B_1 + 1- s$.
Substituting the second into the first provides
\begin{equation*}
\begin{split}
	\left(s+\frac{B_0 - B_1 + 1-s}{2}\right) \frac{B_0 - B_1 + 1- s}{2} =B_0
	\quad \Longleftrightarrow \quad
	s=  \sqrt{(B_0 - B_1 + 1)^2 - 4B_0},
\end{split}
\end{equation*}
under Assumptions \ref{as:neq} and \ref{as:ex}, meaning that $s$ is identified.
Since $s$, $B_0$, and $B_1$ are identified, $m_0$ and $m_1$ are also identified by $m_0= (B_0-B_1+1-s)/2$ and $s=1-m_0-m_1$.
Hence, $p_z^* = E(T^*|Z=z)$ for each $z=0,1$ is identified based on identified parameters $p_z$, $m_0$, and $m_1$, which implies the identification of $\beta^*$.
\begin{flushright}
$\Box$
\end{flushright}

\subsection{Proof of Theorem \ref{thm:gmm}}\label{sec:proof2}
To show the statement, it is sufficient to show that moment condition $E[g(X, \theta_0)]=0$ with $g(X, \theta_0)$ defined in \eqref{eq:moment} is implied by the identification result in Theorem \ref{thm:iden}.

The first moment condition in $E[g(X, \theta_0)]=0$ is $r=E(Z)$ by construction.
The second moment condition in $E[g(X, \theta_0)]=0$ is
\begin{equation*}
\begin{split}
	& E\left[ \Big( m_{0z} + (1-m_{0z}-m_{1z})p_{zv_k}^* - T \Big) I_{zv_k} \right] =0 \\
	&\Longleftrightarrow m_{0z} + (1-m_{0z}-m_{1z})p_{zv_k}^* - p_{zv_k} =0.
\end{split}
\end{equation*}
This equation is equivalent to \eqref{eq:pW}.

We examine the third moment restriction.
By substituting the second moment condition into the third moment condition, we have
\begin{equation*}
\begin{split}
	& E\left[ \left( \tau_z^* + \frac{YT-(1-m_{1z})p^*_{z v_k}\tau_z^*}{m_{0z} + (1-m_{0z}-m_{1z})p^*_{zv_k}} - \frac{Y(1-T)+(1-m_{0z})(1-p^*_{zv_k})\tau_z^*}{1-[m_{0z}+(1-m_{0z}-m_{1z})p^*_{z v_k}]}\right) I_{zv_k}\right] =0\\
	& \Longleftrightarrow E\left[ \left( \tau_z^* + \frac{YT}{p_{zv_k}} - \frac{(1-m_{1z})p^*_{z v_k}\tau_z^*}{p_{zv_k}} - \frac{Y(1-T)}{1-p_{zv_k}} - \frac{(1-m_{0z})(1-p^*_{zv_k})\tau_z^*}{1-p_{zv_k}}\right)I_{z v_k}\right] =0\\
	& \Longleftrightarrow E\left[ \left( \tau_z^* + \frac{YT}{p_{zv_k}} - \frac{(1-m_{1z})\tau_z^*}{p_{zv_k}}\frac{p_{zv_k}-m_{0z}}{1-m_{0z}-m_{1z}} \right.\right.\\
	& \qquad \qquad \left. \left. - \frac{Y(1-T)}{1-p_{zv_k}} - \frac{(1-m_{0z})\tau_z^*}{1-p_{zv_k}}\frac{1-m_{1z}-p_{zv_k}}{1-m_{0z}-m_{1z}}\right)I_{z v_k}\right] =0,\\
\end{split}
\end{equation*}
where the last follows from \eqref{eq:pstar}.
Since $p_{z v_k} = E(TI_{z v_k})/E(I_{z v_k})$, we have
\begin{equation*}
\begin{split}
	& \tau_z^* + \frac{E(YTI_{zv_k})}{E(TI_{z v_k})} - \frac{(1-m_{1z})\tau_z^*}{p_{zv_k}}\frac{p_{zv_k}-m_{0z}}{1-m_{0z}-m_{1z}} \\
	& \qquad \qquad - \frac{E(Y(1-T)I_{zv_k})}{E((1-T)I_{zv_k})} - \frac{(1-m_{0z})\tau_z^*}{1-p_{zv_k}}\frac{1-m_{1z}-p_{zv_k}}{1-m_{0z}-m_{1z}} =0\\
	& \Longleftrightarrow E(Y|T=1,Z=z,V=v_k) - E(Y|T=0,Z=z,V=v_k) \\
	& \qquad \qquad = \left( \frac{1-m_{1z}}{p_{zv_k}}\frac{p_{zv_k}-m_{0z}}{1-m_{0z}-m_{1z}} + \frac{1-m_{0z}}{1-p_{zv_k}}\frac{1-m_{1z}-p_{zv_k}}{1-m_{0z}-m_{1z}}  - 1 \right) \tau_{z}^*\\
	& \Longleftrightarrow E(Y|T=1,Z=z,V=v_k) - E(Y|T=0,Z=z,V=v_k) \\
	& \qquad \qquad = \frac{1}{1-m_{0z}-m_{1z}}\left( 1- \frac{m_{0z}(1-m_{1z})}{p_{zv_k}} - \frac{(1-m_{0z})m_{1z}}{1-p_{zv_k}} \right) \tau_{z}^*,
\end{split}
\end{equation*}
which is identical to \eqref{eq:M} under Assumption \ref{as:ex}.

Noting that $E(TZ)=E(T|Z=1)E(Z)$, the fourth moment condition is rearranged as follows:
\begin{align*}
	& E \left[ \Delta p^* - \left( \frac{TZr^{-1} -m_{01}}{1-m_{01}-m_{11}} - \frac{T(1-Z)(1-r)^{-1} -m_{00}}{1-m_{00}-m_{10}} \right) \right] = 0\\
	& \Longleftrightarrow \Delta p^* - \left(  \frac{p_1 -m_{01}}{1-m_{01}-m_{11}} - \frac{p_0 -m_{00}}{1-m_{00}-m_{10}} \right) = 0,
\end{align*}
which is equal to \eqref{eq:delta_ps}.
Similarly, noting that $E(YZ)=E(Y|Z=1)E(Z)$, the fifth moment condition is
\begin{align*}
	E \left( \beta^* - \frac{YZr^{-1} - Y(1-Z)(1-r)^{-1}}{\Delta p^*} \right) = 0
	\quad \Longleftrightarrow \quad
	\beta^* = \frac{\mu_1 - \mu_0}{p_1^* - p_0^*},
\end{align*}
which is identical to \eqref{eq:taustar}.

Therefore, from the identification result in Theorem \ref{thm:iden}, it holds that $E[g(X, \theta_0)]=0$ with the unique solution $\theta_0$.
\begin{flushright}
	$\Box$
\end{flushright}

\section*{Acknowledgements}
The author is grateful to the editors and two anonymous referees for many valuable comments and suggestions.
The author would also like to thank Ryo Kambayashi, Kengo Kato, Arthur Lewbel, Yoshihiko Nishiyama, Ryo Okui, Takuya Ura, Daniel Wilhelm, Yohei Yamamoto, and the seminar participants of the Economics and Statistics Workshop of Hitotsubashi University, the 2016 Japan-Korea Allied Conference in Econometrics, the 2016 Kansai Econometrics Meeting, and the 2017 Econometric Society European Meeting for their helpful comments and discussions.
The author acknowledges the financial support from the Japan Society for the Promotion of Science under KAKENHI Grant Nos. 15H06214 and 17K13715.

\bibliographystyle{abbrvnat}
\bibliography{ref.bib}

\newpage 

\begin{table}[!p]
\caption{Monte Carlo simulation results for the LATE $\beta^*$} \label{table:monte:LATE}
\begin{center}
\begin{tabular}{rrrcrrrrcrrrr}
\hline\hline
\multicolumn{3}{c}{\bfseries }&\multicolumn{1}{c}{\bfseries }&\multicolumn{4}{c}{\bfseries GMM}&\multicolumn{1}{c}{\bfseries }&\multicolumn{4}{c}{\bfseries IV}\tabularnewline
\cline{5-8} \cline{10-13}
\multicolumn{1}{c}{design}&\multicolumn{1}{c}{$n$}&\multicolumn{1}{c}{true}&\multicolumn{1}{c}{}&\multicolumn{1}{c}{bias}&\multicolumn{1}{c}{SD}&\multicolumn{1}{c}{RMSE}&\multicolumn{1}{c}{CP}&\multicolumn{1}{c}{}&\multicolumn{1}{c}{bias}&\multicolumn{1}{c}{SD}&\multicolumn{1}{c}{RMSE}&\multicolumn{1}{c}{CP}\tabularnewline
\hline
$1$&$ 200$&$1.000$&&$0.336$&$0.834$&$0.899$&$0.966$&&$1.335$&$ 4.508$&$ 4.701$&$0.990$\tabularnewline
&$ 500$&$1.000$&&$0.180$&$0.540$&$0.569$&$0.942$&&$1.156$&$ 1.082$&$ 1.583$&$0.764$\tabularnewline
&$1000$&$1.000$&&$0.090$&$0.333$&$0.345$&$0.949$&&$1.052$&$ 0.478$&$ 1.156$&$0.261$\tabularnewline
\hline
$2$&$ 200$&$1.000$&&$0.389$&$1.294$&$1.351$&$0.963$&&$1.662$&$12.171$&$12.284$&$0.995$\tabularnewline
&$ 500$&$1.000$&&$0.244$&$0.639$&$0.684$&$0.957$&&$1.154$&$ 0.958$&$ 1.500$&$0.881$\tabularnewline
&$1000$&$1.000$&&$0.144$&$0.456$&$0.479$&$0.943$&&$1.041$&$ 0.562$&$ 1.183$&$0.561$\tabularnewline
\hline 
$3$&$ 200$&$1.000$&&$0.345$&$0.811$&$0.881$&$0.967$&&$1.729$&$29.042$&$29.094$&$0.989$\tabularnewline
&$ 500$&$1.000$&&$0.171$&$0.505$&$0.534$&$0.953$&&$1.179$&$ 1.453$&$ 1.871$&$0.764$\tabularnewline
&$1000$&$1.000$&&$0.098$&$0.339$&$0.353$&$0.951$&&$1.047$&$ 0.460$&$ 1.144$&$0.215$\tabularnewline
\hline 
$4$&$ 200$&$1.000$&&$0.292$&$1.092$&$1.130$&$0.952$&&$1.348$&$ 4.216$&$ 4.427$&$0.991$\tabularnewline
&$ 500$&$1.000$&&$0.233$&$0.685$&$0.723$&$0.947$&&$1.132$&$ 1.289$&$ 1.716$&$0.894$\tabularnewline
&$1000$&$1.000$&&$0.157$&$0.436$&$0.463$&$0.955$&&$1.045$&$ 0.577$&$ 1.193$&$0.555$\tabularnewline
\hline 
$5$&$ 200$&$1.000$&&$0.294$&$0.706$&$0.765$&$0.965$&&$1.122$&$ 5.389$&$ 5.504$&$0.974$\tabularnewline
&$ 500$&$1.000$&&$0.174$&$0.458$&$0.490$&$0.951$&&$1.119$&$ 0.633$&$ 1.286$&$0.629$\tabularnewline
&$1000$&$1.000$&&$0.103$&$0.324$&$0.339$&$0.946$&&$1.062$&$ 0.417$&$ 1.141$&$0.107$\tabularnewline
\hline 
$6$&$ 200$&$1.000$&&$0.244$&$0.989$&$1.019$&$0.954$&&$1.235$&$ 3.061$&$ 3.301$&$0.989$\tabularnewline
&$ 500$&$1.000$&&$0.213$&$0.583$&$0.621$&$0.949$&&$1.094$&$ 0.792$&$ 1.350$&$0.810$\tabularnewline
&$1000$&$1.000$&&$0.162$&$0.417$&$0.448$&$0.957$&&$1.026$&$ 0.490$&$ 1.137$&$0.419$\tabularnewline
\hline
\end{tabular}\end{center}
\end{table}

\begin{table}[!p]
\caption{Monte Carlo simulation results for the first-stage regression $\Delta p^*$} \label{table:monte:first}
\begin{center}
\begin{tabular}{rrrcrrrrcrrrr}
\hline\hline
\multicolumn{3}{c}{\bfseries }&\multicolumn{1}{c}{\bfseries }&\multicolumn{4}{c}{\bfseries GMM}&\multicolumn{1}{c}{\bfseries }&\multicolumn{4}{c}{\bfseries OLS}\tabularnewline
\cline{5-8} \cline{10-13}
\multicolumn{1}{c}{design}&\multicolumn{1}{c}{$n$}&\multicolumn{1}{c}{true}&\multicolumn{1}{c}{}&\multicolumn{1}{c}{bias}&\multicolumn{1}{c}{SD}&\multicolumn{1}{c}{RMSE}&\multicolumn{1}{c}{CP}&\multicolumn{1}{c}{}&\multicolumn{1}{c}{bias}&\multicolumn{1}{c}{SD}&\multicolumn{1}{c}{RMSE}&\multicolumn{1}{c}{CP}\tabularnewline
\hline
$1$&$ 200$&$0.341$&&$-0.032$&$0.133$&$0.137$&$0.994$&&$-0.168$&$0.069$&$0.182$&$0.326$\tabularnewline
&$ 500$&$0.341$&&$-0.017$&$0.106$&$0.107$&$0.985$&&$-0.170$&$0.044$&$0.176$&$0.029$\tabularnewline
&$1000$&$0.341$&&$-0.011$&$0.080$&$0.081$&$0.979$&&$-0.171$&$0.031$&$0.174$&$0.000$\tabularnewline
\hline 
$2$&$ 200$&$0.341$&&$-0.023$&$0.152$&$0.154$&$0.992$&&$-0.168$&$0.070$&$0.182$&$0.328$\tabularnewline
&$ 500$&$0.341$&&$-0.025$&$0.117$&$0.119$&$0.988$&&$-0.170$&$0.044$&$0.176$&$0.030$\tabularnewline
&$1000$&$0.341$&&$-0.016$&$0.092$&$0.093$&$0.982$&&$-0.169$&$0.031$&$0.172$&$0.000$\tabularnewline
\hline 
$3$&$ 200$&$0.341$&&$-0.033$&$0.131$&$0.135$&$0.995$&&$-0.169$&$0.070$&$0.183$&$0.326$\tabularnewline
&$ 500$&$0.341$&&$-0.015$&$0.104$&$0.105$&$0.982$&&$-0.171$&$0.043$&$0.177$&$0.024$\tabularnewline
&$1000$&$0.341$&&$-0.010$&$0.082$&$0.083$&$0.973$&&$-0.169$&$0.031$&$0.172$&$0.001$\tabularnewline
\hline 
$4$&$ 200$&$0.341$&&$-0.006$&$0.155$&$0.155$&$0.994$&&$-0.165$&$0.068$&$0.178$&$0.339$\tabularnewline
&$ 500$&$0.341$&&$-0.022$&$0.123$&$0.124$&$0.990$&&$-0.170$&$0.044$&$0.175$&$0.034$\tabularnewline
&$1000$&$0.341$&&$-0.020$&$0.095$&$0.097$&$0.983$&&$-0.169$&$0.031$&$0.172$&$0.000$\tabularnewline
\hline 
$5$&$ 200$&$0.382$&&$-0.034$&$0.138$&$0.142$&$0.993$&&$-0.188$&$0.067$&$0.200$&$0.212$\tabularnewline
&$ 500$&$0.382$&&$-0.023$&$0.112$&$0.115$&$0.988$&&$-0.193$&$0.044$&$0.198$&$0.005$\tabularnewline
&$1000$&$0.382$&&$-0.012$&$0.086$&$0.087$&$0.974$&&$-0.191$&$0.032$&$0.194$&$0.000$\tabularnewline
\hline 
$6$&$ 200$&$0.382$&&$-0.013$&$0.156$&$0.157$&$0.994$&&$-0.187$&$0.068$&$0.199$&$0.231$\tabularnewline
&$ 500$&$0.382$&&$-0.025$&$0.125$&$0.128$&$0.984$&&$-0.190$&$0.043$&$0.195$&$0.009$\tabularnewline
&$1000$&$0.382$&&$-0.026$&$0.100$&$0.104$&$0.979$&&$-0.189$&$0.031$&$0.192$&$0.000$\tabularnewline
\hline
\end{tabular}\end{center}
\end{table}

\begin{table}[!p]
\caption{Monte Carlo simulation results for the misclassification probabilities $m_0$ and $m_1$} \label{table:monte:m0}
\begin{center}
\begin{tabular}{rrrcrrrrcrrrr}
\hline\hline
\multicolumn{3}{c}{\bfseries }&\multicolumn{1}{c}{\bfseries }&\multicolumn{4}{c}{\bfseries parameter: $m_0$}&\multicolumn{1}{c}{\bfseries }&\multicolumn{4}{c}{\bfseries parameter: $m_1$}\tabularnewline
\cline{5-8} \cline{10-13}
\multicolumn{1}{c}{design}&\multicolumn{1}{c}{$n$}&\multicolumn{1}{c}{true}&\multicolumn{1}{c}{}&\multicolumn{1}{c}{bias}&\multicolumn{1}{c}{SD}&\multicolumn{1}{c}{RMSE}&\multicolumn{1}{c}{CP}&\multicolumn{1}{c}{}&\multicolumn{1}{c}{bias}&\multicolumn{1}{c}{SD}&\multicolumn{1}{c}{RMSE}&\multicolumn{1}{c}{CP}\tabularnewline
\hline
$1$&$ 200$&$0.250$&&$-0.040$&$0.131$&$0.137$&$0.983$&&$-0.058$&$0.135$&$0.147$&$0.983$
\tabularnewline
&$ 500$&$0.250$&&$-0.031$&$0.110$&$0.114$&$0.958$&&$-0.034$&$0.113$&$0.118$&$0.957$
\tabularnewline
&$1000$&$0.250$&&$-0.020$&$0.087$&$0.090$&$0.943$&&$-0.019$&$0.084$&$0.086$&$0.950$
\tabularnewline \hline
$2$&$ 200$&$0.250$&&$-0.012$&$0.144$&$0.145$&$0.961$&&$-0.079$&$0.146$&$0.166$&$0.984$
\tabularnewline
&$ 500$&$0.250$&&$-0.026$&$0.121$&$0.124$&$0.943$&&$-0.063$&$0.131$&$0.145$&$0.976$
\tabularnewline
&$1000$&$0.250$&&$-0.030$&$0.099$&$0.103$&$0.948$&&$-0.034$&$0.110$&$0.115$&$0.950$
\tabularnewline \hline
$3$&$ 200$&$0.250$&&$-0.033$&$0.130$&$0.135$&$0.981$&&$-0.059$&$0.136$&$0.149$&$0.986$
\tabularnewline
&$ 500$&$0.250$&&$-0.028$&$0.106$&$0.110$&$0.955$&&$-0.029$&$0.108$&$0.112$&$0.958$
\tabularnewline
&$1000$&$0.250$&&$-0.021$&$0.084$&$0.086$&$0.955$&&$-0.022$&$0.086$&$0.088$&$0.953$
\tabularnewline \hline
$4$&$ 200$&$0.250$&&$ 0.005$&$0.133$&$0.133$&$0.951$&&$-0.081$&$0.148$&$0.169$&$0.980$
\tabularnewline
&$ 500$&$0.250$&&$-0.024$&$0.122$&$0.125$&$0.943$&&$-0.066$&$0.135$&$0.150$&$0.966$
\tabularnewline
&$1000$&$0.250$&&$-0.035$&$0.104$&$0.109$&$0.948$&&$-0.040$&$0.112$&$0.119$&$0.951$
\tabularnewline \hline
$5$&$ 200$&$0.250$&&$-0.038$&$0.128$&$0.134$&$0.986$&&$-0.060$&$0.136$&$0.149$&$0.991$
\tabularnewline
&$ 500$&$0.250$&&$-0.032$&$0.111$&$0.116$&$0.955$&&$-0.035$&$0.113$&$0.118$&$0.955$
\tabularnewline
&$1000$&$0.250$&&$-0.021$&$0.084$&$0.087$&$0.948$&&$-0.020$&$0.085$&$0.087$&$0.948$
\tabularnewline \hline
$6$&$ 200$&$0.250$&&$ 0.007$&$0.127$&$0.127$&$0.959$&&$-0.082$&$0.148$&$0.169$&$0.979$
\tabularnewline
&$ 500$&$0.250$&&$-0.025$&$0.118$&$0.121$&$0.939$&&$-0.064$&$0.131$&$0.146$&$0.956$
\tabularnewline
&$1000$&$0.250$&&$-0.033$&$0.102$&$0.107$&$0.938$&&$-0.047$&$0.114$&$0.123$&$0.953$
\tabularnewline
\hline
\end{tabular}\end{center}
\end{table}

\begin{table}[!p]
\caption{Descriptive statistics in the empirical illustration with 3,010 observations} \label{table:empirical:descriptive}
\begin{center}
\begin{tabular}{lccccc}
\hline\hline
 &&\multicolumn{1}{c}{mean}&\multicolumn{1}{c}{SD}&\multicolumn{1}{c}{min}&\multicolumn{1}{c}{max}\tabularnewline
\hline
$lwage$&&$6.262$&$0.444$&$4.605$&$7.785$\tabularnewline
$college$&&$0.412$&$0.492$&$0$&$1$\tabularnewline
$proximity4$&&$0.682$&$0.466$&$0$&$1$\tabularnewline
$proximity2$&&$0.441$&$0.497$&$0$&$1$\tabularnewline
$proximity4 \cdot proximity2$&&$0.328$&$0.470$&$0$&$1$\tabularnewline
$(1 - proximity4) \cdot proximity2$&&$0.113$&$0.316$&$0$&$1$\tabularnewline
$proximity4 \cdot (1 - proximity2)$&&$0.354$&$0.478$&$0$&$1$\tabularnewline
$(1 - proximity4) \cdot (1 - proximity2)$&&$0.205$&$0.404$&$0$&$1$\tabularnewline
\hline
\end{tabular}\end{center}
\end{table}

\begin{landscape}
\begin{table}[!p]
\caption{Estimation ignoring the measurement error} \label{table:empirical:ign}
\begin{center}

\begin{tabular}{lcccccccc}
\hline\hline
\multicolumn{1}{l}{estimation}&&\multicolumn{1}{c}{(1) IV}&\multicolumn{1}{c}{(2) IV}&\multicolumn{1}{c}{(3) OLS}&\multicolumn{1}{c}{(4) OLS}&\multicolumn{1}{c}{(5) OLS}&\multicolumn{1}{c}{(6) OLS}&\multicolumn{1}{c}{(7) OLS}\tabularnewline
\hline
data&&&&&&&$proximity4 = 1$&$proximity4 = 0$\tabularnewline
\hline
dependent variable&&$lwage$&$lwage$&$lwage$&$college$&$college$&$college$&$college$\tabularnewline
\hline
$college$&&1.317&1.790&0.198&&&&\tabularnewline
&&(0.227)&(0.672)&(0.016)&&&&\tabularnewline
$proximity4$&&&&&0.118&&&\tabularnewline
&&&&&(0.019)&&&\tabularnewline
$proximity2$&&&&&&0.049&0.076&$-0.056$\tabularnewline
&&&&&&(0.018)&(0.022)&(0.032)\tabularnewline
\hline
instrument&&$proximity4$&$proximity2$&&&&&\tabularnewline
\hline
observations&&3010&3010&3010&3010&3010&2053&957\tabularnewline
\hline
\end{tabular}

\begin{tablenotes}
Note: Each column gives the naive estimate ignoring the measurement error based on each estimation procedure.
The numbers in the parentheses are heteroscedasticity robust standard errors.
\end{tablenotes}

\end{center}
\end{table}
\end{landscape}

\begin{landscape}

\begin{table}[!p]
\caption{GMM estimation result with four-year college proximity $Z$ and two-year college proximity $V$} \label{table:empirical:gmm1}
\begin{center}
\begin{tabular}{ccccccccccc}
\hline\hline
$\beta^*$&$\Delta p^*$&$r$&$m_0$&$m_1$&$p_{00}^*$&$p_{01}^*$&$p_{10}^*$&$p_{11}^*$&$\tau_0^*$&$\tau_1^*$\tabularnewline
\hline
0.421&0.393&0.682&0.296&0.402&0.191&0.021&0.387&0.642&0.599&0.588\tabularnewline
(0.124)&(0.110)&(0.008)&(0.033)&(0.069)&(0.104)&(0.074)&(0.092)&(0.131)&(0.370)&(0.205)\tabularnewline
\hline
\end{tabular}

\begin{tablenotes}
Note: The table provides the estimates and heteroscedasticity robust standard errors for the GMM estimation based on the identity weighting matrix.
This is the main result in the empirical illustration.
The number of observations is 3,010.
The parameters are the LATE $\beta^*$, true first-stage regression $\Delta p^*$, mean $r = E(Z)$, misclassification probabilities $m_0$ and $m_1$, true conditional treatment probability $p_{zv}^* = \Pr(T^*=1|Z=z, V=v)$, and $\tau_z^* = E(Y|T^*=1, Z=z) - E(Y|T^*=0, Z=z)$.
\end{tablenotes}

\end{center}
\end{table}

\begin{table}[!p]
\caption{GMM estimation result with two-year college proximity $Z$ and four-year college proximity $V$} \label{table:empirical:gmm2}
\begin{center}
\begin{tabular}{ccccccccccc}
\hline\hline
$\beta^*$&$\Delta p^*$&$r$&$m_0$&$m_1$&$p_{00}^*$&$p_{01}^*$&$p_{10}^*$&$p_{11}^*$&$\tau_0^*$&$\tau_1^*$\tabularnewline
\hline
0.441&0.179&0.441&0.254&0.475&0.365&0.591&0.152&0.870&0.507&1.372\tabularnewline
(0.806)&(0.323)&(0.009)&(0.276)&(0.239)&(0.326)&(0.136)&(0.755)&(0.627)&(1.067)&(8.234)\tabularnewline
\hline
\end{tabular}

\begin{tablenotes}
Note: The table provides the estimates and heteroscedasticity robust standard errors for the GMM estimation based on the identity weighting matrix.
This is an auxiliary result in the empirical illustration.
The number of observations is 3,010.
The parameters are the LATE $\beta^*$, true first-stage regression $\Delta p^*$, mean $r = E(Z)$, misclassification probabilities $m_0$ and $m_1$, true conditional treatment probability $p_{zv}^* = \Pr(T^*=1|Z=z, V=v)$, and $\tau_z^* = E(Y|T^*=1, Z=z) - E(Y|T^*=0, Z=z)$.
\end{tablenotes}

\end{center}
\end{table}

\end{landscape}

\end{document}